\documentclass[conference]{IEEEtran}
\IEEEoverridecommandlockouts
\usepackage{mathptmx} 
\usepackage{fancyhdr}
\usepackage[normalem]{ulem}
\usepackage[hyphens]{url}
\usepackage[sort,nocompress]{cite}
\usepackage[final]{microtype}
\usepackage[keeplastbox]{flushend}
\usepackage[bookmarks=true,breaklinks=true,letterpaper=true,colorlinks,linkcolor=black,citecolor=blue,urlcolor=black]{hyperref}

\usepackage{subcaption}
\usepackage{cite}
\usepackage{amsmath,amssymb,amsfonts}
\usepackage{algorithmic}
\usepackage{graphicx}
\usepackage{enumitem}
\setlist{nosep}
\usepackage{setspace}
\usepackage{graphics}
\usepackage{float}
\usepackage{dblfloatfix}    
\usepackage{rotating}
\usepackage{tikz}
\usepackage{textcomp}
\usepackage{xcolor}
\usepackage{fancyhdr}
\usepackage[hyphens]{url}
\usepackage{makecell}
\usepackage{tabularx}
\usepackage[capitalise]{cleveref}
\usepackage{url}

\usepackage{xspace}
\usepackage{siunitx}
\usepackage{fancyhdr}
\usepackage{lastpage}
\usepackage{amsmath}
\usepackage{blindtext}
\usepackage{comment}
\usepackage[english]{babel}
\usepackage{blindtext}

\def\BibTeX{{\rm B\kern-.05em{\sc i\kern-.025em b}\kern-.08em
    T\kern-.1667em\lower.7ex\hbox{E}\kern-.125emX}}

\fancypagestyle{firstpage}{
  \fancyhf{}
  
  \fancyfoot[C]{\thepage}
}
\title{Guidelines for Submission to MICRO 2020} 
\usepackage{listings}

\newcommand{\OURLCORE}{FPRaker} 
\newcommand{\OURSCORE}{FPRaker}
\newcommand{\OURL}{\textit{\OURLCORE}\xspace} 
\newcommand{\OURS}{\textit{\OURSCORE}\xspace} 

\usepackage{lipsum}

\usepackage{array}
\newcolumntype{C}[1]{>{\centering\let\newline\\\arraybackslash\hspace{0pt}}m{#1}}

\makeatletter
\AtBeginDocument{\let\hl\@firstofone}
\makeatother

\begin{document}

\title{\OURLCORE:  A Processing Element For Accelerating Neural Network Training}



%
\author{\IEEEauthorblockN{Omar Mohamed Awad\IEEEauthorrefmark{1},
Mostafa Mahmoud\IEEEauthorrefmark{1},
Isak Edo\IEEEauthorrefmark{1},
Ali Hadi Zadeh\IEEEauthorrefmark{1},
Ciaran Bannon\IEEEauthorrefmark{1},
Anand Jayarajan\IEEEauthorrefmark{1}\\
Gennady Pekhimenko\IEEEauthorrefmark{1,2},
Andreas Moshovos\IEEEauthorrefmark{1,2}\\
\{omar.awad, mostafa.mahmoud, isak.edo, a.hadizadeh, ciaran.bannon\}@mail.utoronto.ca,\\
\{anandj, pekhimenko\}@cs.toronto.edu, moshovos@ece.utoronto.ca
}
\IEEEauthorblockA{\IEEEauthorrefmark{1} University of Toronto, \IEEEauthorrefmark{2} Vector Institute}}

\maketitle
\thispagestyle{firstpage}
\pagestyle{plain}
\footnotetext[1]{This version of the paper was submitted to MICRO2020. A previous version was submitted to ISCA2020.}%

\begin{abstract}
We present \OURL, a processing element for composing training accelerators. \OURL processes several floating-point multiply-accumulation operations concurrently and accumulates their result into a higher precision accumulator. \OURL boosts performance and energy efficiency during training by taking advantage of the values that naturally appear during training. It processes the significand of the operands of each multiply-accumulate as a series of signed powers of two. The conversion to this form is done on-the-fly. This exposes ineffectual work that can be skipped: values when encoded have few terms and some of them can be discarded as they would fall outside the range of the accumulator given the limited precision of floating-point. \OURL also takes advantage of spatial correlation in values across channels and uses delta-encoding  off-chip to reduce memory footprint and bandwidth. We demonstrate that \OURL can be used to compose an accelerator for training and that it can improve performance and energy efficiency compared to using optimized bit-parallel floating-point units under iso-compute area constraints. We also demonstrate that \OURL delivers additional benefits when training incorporates pruning and quantization. Finally, we show that \OURL naturally amplifies performance with training methods that use a different precision per layer.
\end{abstract}
\vspace*{-0.2cm}

\section{Motivation}\label{motiv}

The pervasive applications of deep learning and the end of Dennard scaling have been driving efforts for accelerating deep learning inference and training. These efforts span the full system stack, from algorithms, to middleware and hardware architectures. Here we target training, a task that includes inference as a subtask. Training is a compute- and memory-intensive task often requiring weeks of compute time even with state-of-the-art training methods and hardware~\cite{ScaleDeep}. The cost of training is staggering. For example, training XLNetrequires 512 TPU v3 chips for 2.5 days leading to a total training cost above \$61K~\cite{XLNet}. Further, training Grover-Mega requires 128 TPU v3 chips for 2 weeks leading to a total training cost of \$25K~\cite{GroverMega}. Often multiple trials are required for hyperparameter tuning which further amplifies the cost.

During training a set of annotated inputs, that is inputs for which the desired output is known is processed by repeatedly performing a forward and backward pass. The \textit{forward} pass performs inference whose output is initially inaccurate. However, given that the desired outputs are known, the training algorithm can calculate a \textit{loss}, a metric of how far the outputs are from the desired ones. During the \textit{backward} pass,  this loss is used to adjust the network's parameters and to have it slowly converge to its best possible accuracy.

Numerous methods have been developed to accelerate training, and fortunately often they can be used in combination.  Distributed training partitions the training workload across several computing nodes taking advantage of data, model, or pipeline parallelism~\cite{shallue2019measuring, pipedream, gpipe, megatron,hypar}. Timing communication and computation can further reduce training time~\cite{Wen:2017,UCNN,hashemi2018tictac,GenericComm,jayarajan2019priority}. Dataflow optimizations to facilitate data blocking and to maximize data reuse reduces the cost of on- and off-chip accesses within the node maximizing reuse from lower cost components of the memory hierarchy~\cite{EyerissISCA2016,Gao:Tetris:ASPLOS}. Another family of methods reduces the footprint of the intermediate data needed during training. For example, in the simplest form of training, all neuron values produced during the forward pass are kept to be used during backpropagation. Batching and keeping only one or a few samples instead reduces this cost. Lossless and lossy compression methods further reduce the footprint of such data~\cite{rhu2018compressing,GIST,LIN2018,Wen:2017,alistarh2017qsgd,seide2014-bit}. 
Finally, selective backpropagation methods alter the backward pass by propagating loss only for some of the neurons~\cite{MeProp} thus reducing work. 

On the other hand, the need to boost energy efficiency during inference has led to techniques that increase computation and memory needs during training. This includes works that perform network pruning and quantization during training.  Pruning zeroes out weights and thus creates an opportunity for reducing work and model size during inference.  Quantization produces models that use shorter and more energy efficient to compute with datatypes such as 16b, 8b or 4b fixed-point values. Parameter Efficient Training~\cite{dynamic_sparse_reparam}, Memorized Sparse Back-propagation~\cite{mem_sparse_backprop} are examples of recent pruning methods. PACT~\cite{PACT} and outlier-aware quantization~\cite{outlier} are training time quantization methods. Network architecture search techniques also increase training time as they adjust the model's architecture~\cite{elsken2018neural}.

Despite the successes already reported via the aforementioned methods, and while we expect that these methods will be refined further, the need to further accelerate training both at the data center and at the edge remains unabated. Operating and maintenance costs, latency, throughput, and node count are major considerations for data centers. At the edge energy and latency are major considerations where training may be primarily used to refine or augment already trained models~\cite{DBLP:conf/mm/XiaoZYPZ14, DBLP:conf/eccv/CastroMGSA18, DBLP:journals/corr/abs-1803-10232}. Regardless of the target application, improving node performance would be of value. Accordingly, in this work we target methods that could \textit{complement} existing training acceleration methods. In general, the bulk of the computations and data transfers during training is for performing  multiply-accumulate operations (MAC) during the forward and backward passes. As mentioned above, compression methods can greatly reduce the cost of data transfers. Here we target the processing elements for these operations and propose designs that exploit \textit{ineffectual} work that occurs naturally during training and whose frequency is amplified by quantization, pruning, and selective backpropagation. 

To devise a processing element that can eliminate ineffectual work during training we may attempt to build upon the numerous proposals that exploit ineffectual work during \textit{inference}. We highlight some those approaches that are most relevant to this work. Any MAC operation where any of the two input operands, be it the activation or the weight is ineffectual. The first class of accelerators rely on that zeros occur naturally in the activations of many models especially when they use ReLU~\cite{eyeriss,han_eie:isca_2016,Cnvlutin,ZeNa}.  There are several accelerators that target pruned models e.g.,~\cite{han_eie:isca_2016,cambricon:2016,SCNN_ISCA,CambriconXMICRO16,CambriconS:MICRO2018,pragmatic,DBLP:conf/asplos/KungMZ19,MarsSparse,Sparten,DBLP:journals/esticas/ChenYES19}. Another class of designs benefit from reduced value ranges whether these occur naturally or result from quantization. This includes bit-serial designs~\cite{Stripes-MICRO,DBLP:conf/isca/EckertWWSISBD18,LOOM,DBLP:journals/jssc/LeeKKSKY19, Shapeshifter}, and designs that support many different datatypes such as BitFusion~\cite{bitfusion}. Finally, another class of designs targets \textit{bit-sparsity} where by decomposing multiplication into a series of shift-and-add operations they expose ineffectual work at the bit-level~\cite{pragmatic,Tactical,laconic}.

While we can draw from the experience gained from the aforementioned design for inference, training presents us with different challenges. First, is the \textit{datatype}. While models during inference work with fixed-point values of relatively limited range, the values training operates upon tend to be spread over a large range. Accordingly, training implementations use floating-point arithmetic with single-precision IEEE floating point arithmetic (FP32) being sufficient for virtually all models. Other datatypes that facilitate the use of more energy- and area-efficient multiply-accumulate units compared to FP32 have been successfully used in training many models. These include bfloat16, and 8b or smaller floating-point formats~\cite{gupta2015deep,bfloat16Google,intel_bf16,narrowFP_ibm,DBLP:conf/iclr/0002MMKAB0VKGHD18,Koster:2017:FAN:3294771.3294937, HALP}. Moreover, since floating-point arithmetic is a lot more expensive than integer arithmetic, mixed datatype training methods use floating-point arithmetic only sparingly~\cite{mixedP, DBLP:conf/iclr/0002MMKAB0VKGHD18, Drumond:2018:TDH:3326943.3326985,  nvidia_mixedP}. Despite these proposals, FP32 remains today the standard fall-back format, especially for training on large and challenging datasets. As a result of its limited range and the lack of an exponent, the fixed-point representation used during inference gives rise to zero values (too small a value to be represented), zero bit prefixes (small value that can be represented), and bit sparsity (most values tend to be small and few are large) that the aforementioned inference accelerators rely upon. FP32 can represent much smaller values, its mantissa is normalized, and whether bit sparsity exists has not been demonstrated.

Second, is the \textit{computation structure}. Inference operates on two tensors, the weights and the activations, performing per layer a matrix/matrix or matrix/vector multiplication or pairwise vector operations to produce the activations for the next layer in a feed-forward fashion. Training includes this computation as its \textit{forward} pass which is followed by the \textit{backward} pass that involves a third tensor, the gradients. Most importantly, the backward pass uses the activation and weight tensors in a different way than the forward pass, making it difficult to pack them efficiently in memory, more so to remove zeros as done by inference accelerators that target sparsity. Related to computation structure, third, is \textit{value mutability} and \textit{value content}. Whereas in inference the weights are static, they are not so during training. Furthermore, training initializes the network with random values which it then slowly adjusts. Accordingly, one cannot necessarily expect the values processed during training to exhibit similar behavior such as sparsity or bit-sparsity. More so, for the gradients which are values that do not appear at all during inference.

\textbf{Our first contribution} is that we demonstrate that a large fraction of the work performed during training is ineffectual. To expose this ineffectual work we decompose each multiplication into a series of single bit multiply-accumulate operations. This reveals two sources of ineffectual work: First, more than 85\%\ of the computations are ineffectual since one of the inputs is zero. Second, the combination of the high dynamic range (exponent) and the limited precision (mantissa) often yields values which are non-zero, yet too small to affect the accumulated result, even when using extended precision (e.g., trying to accumulate $2^{-64}$ into $2^{64}$). 

This observation led us to consider whether it is possible to use \textit{bit-skipping} (bit-serial where we skip over zero bits) processing to exploit these two behaviors. Bit-skipping processing of multiply-accumulate operations has been proposed before for inference. Bit-Pragmatic is a data-parallel processing element that performs such bit-skipping of one operand side~\cite{pragmatic} whereas Laconic does so for both sides~\cite{laconic}. Since these methods target inference only they work with fixed-point values. Since we found that there is little bit-sparsity in the weights during training, we focused on Bit-Pragmatic's approach. Converting a fixed-point design to floating-point is a non-trivial task to start with. Regardless, converting Bit-Pragmatic into floating-point resulted in an area-expensive unit which performs poorly under iso-compute area constraints. Specifically, compared to an \textit{optimized} Bfloat16 processing element (see Section~\ref{sec:methodology}) that performs 8 MAC operations, under iso-compute constraints, an optimized accelerator configuration using the Bfloat16 Bit-Pragmatic PEs is on average $1.72\times$ slower and $1.96\times$ less energy efficient. In the worst case, the Bfloat16 bit-pragmatic PE was $2.86\times$ slower and $3.2\times$ less energy efficient. The Bfloat16 Bit-Pragmatic PE is $2.5\times$ smaller than the bit-parallel PE, and while we can use more such PEs for the same area, we cannot fit enough of them to boost performance via parallelism as required by all bit-serial and bit-skipping designs.

Accordingly, \textbf{our second contribution} is \OURL, a processing tile for \textit{training} accelerators which exploits both bit-sparsity and out-of-bounds computations. \OURL comprises several adder-tree based processing elements organized in a grid so that it can exploit data reuse both spatially and temporally. The processing elements multiply multiple value pairs concurrently and accumulate their products into an output accumulator. They process one of the input operands per multiplication as a series of signed powers of two hitherto referred to as \textit{terms}. The conversion of that operand into powers of two is performed on the fly; all operands are stored in floating point form in memory. The processing elements take advantage of ineffectual work that stems either from mantissa bits that were zero or from out-of-bounds multiplications given the current accumulator value.  The tile is designed for area efficiency and it incorporates the following designs choices for this purpose: a)~The processing element limits the range of powers-of-two that they can be processed simultaneously greatly reducing the cost of its shift-and-add components. b)~A common exponent processing unit that is time-multiplexed among multiple processing elements. c)~The power-of-two encoders are shared along the rows.  d)~Per processing element buffers reduce the effects of work imbalance across the processing elements. e)~The PE implements a low cost mechanism for eliminating out-of-range intermediate values. Skipping out-of-bound intermediate values not only reduces the amount of work, but more importantly proves very effective in reducing the effect of cross-lane synchronization. Section~\ref{FP-PRA} explains that a bit-parallel unit could only take advantage of these values to power-gate some of its component but not to also improve performance.

In more detail, \OURL has the following characteristics:
a)~It does not affect numerical accuracy. The results it produces adhere to the floating-point arithmetic used during training. b)~It skips ineffectual operations that would result from zero mantissa bits and from out-of-range intermediate values. c)~Despite  individual MAC\ operations taking more than one cycle, \OURL's computational throughput is higher compared to conventional floating-point units; given that \OURL processing elements are much smaller we can fit more of them in the same area. c)~It naturally supports shorter mantissa lengths thus rewarding innovation in training methods with mixed or shorter datatypes~\cite{narrowFP_ibm,sakr2019accumulation}. It does so while not requiring that the methods be universally applicable to all models. d)~\OURL allows us to choose which tensor input we wish to process serially per layer. This allows us to target those tensors that have more sparsity depending on the layer and the pass (forward or backward).

\textbf{Our third contribution} is a simple, low-overhead memory encoding for floating-point values that relies on the value distribution that is typical in deep learning training. We observed that consecutive values across channels have similar values and thus exponents. Accordingly, we encode the exponents as deltas for groups of such values. We use this encoding when storing and reading values off chip, thus further reducing the cost of memory transfers.


\sloppy
We highlight the following experimental observations:
a)~While some neural networks naturally exhibit zero values (sparsity) during training, unless pruning is used, this is limited to the activations and the gradients. b)~term-sparsity exists in all tensors including the weights and is much higher than sparsity. 
c)~Compared to an accelerator using optimized bit-parallel FP32 processing elements and that can perform 4K bfloat16~\cite{bfloat16Google,intel_bf16} MACs per cycle, a configuration that uses the same compute area to deploy \OURL PEs is $1.5\times$ faster and $1.4\times$ more energy efficient.
d)~Performance benefits with \OURL remain fairly stable throughout the training process for all three major operations.
e)~\OURL can be used in conjunction with training methods that specify a different accumulator precision to be used per layer. There it can improve performance vs using an accumulator with a fixed width significand by 38\% for ResNet18.

\begin{figure*}[h!]
\begin{minipage}{.7\textwidth}
  \subfloat[Value Sparsity]{\includegraphics[scale=0.19]{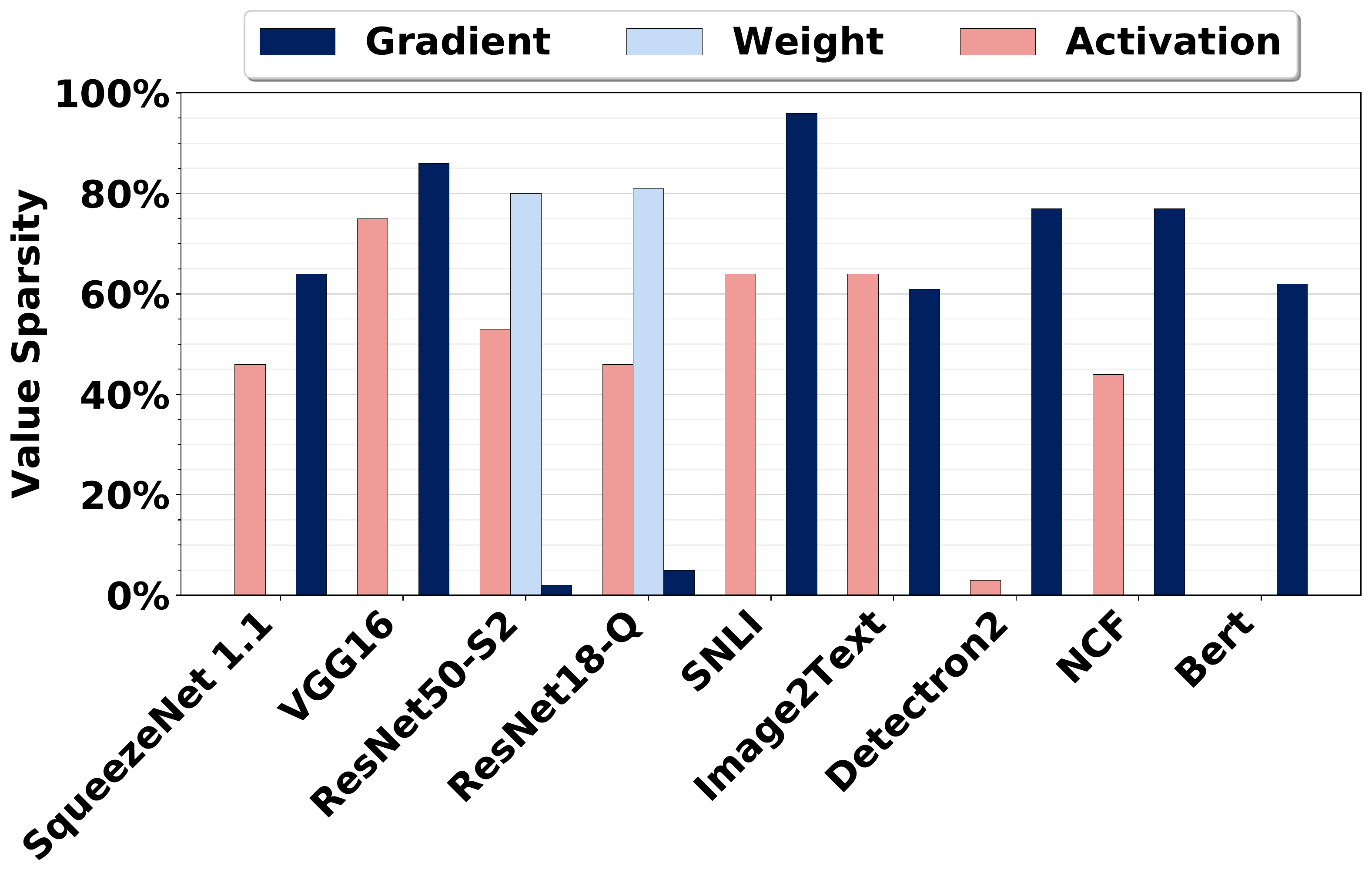}\label{fig:value_sparsity}}
  \hfill
  \subfloat[Term Sparsity]{\includegraphics[scale=0.19]{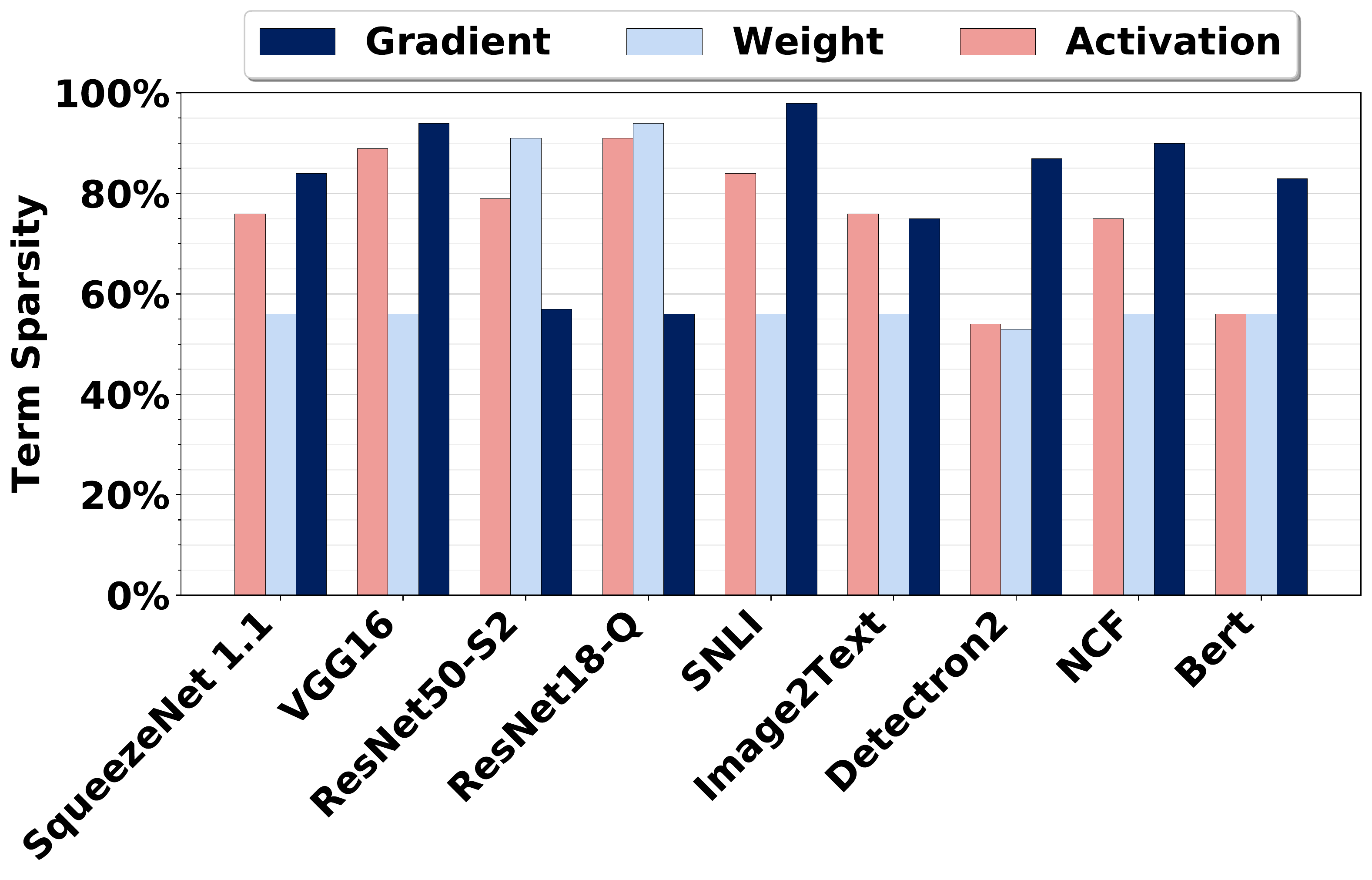}\label{fig:term_sparsity}}
  \vspace*{-0.2cm}
  \caption{Value and Term Sparsity in Tensors During Training.}
\end{minipage}%
\begin{minipage}{.3\textwidth}
  \includegraphics[scale=0.19]{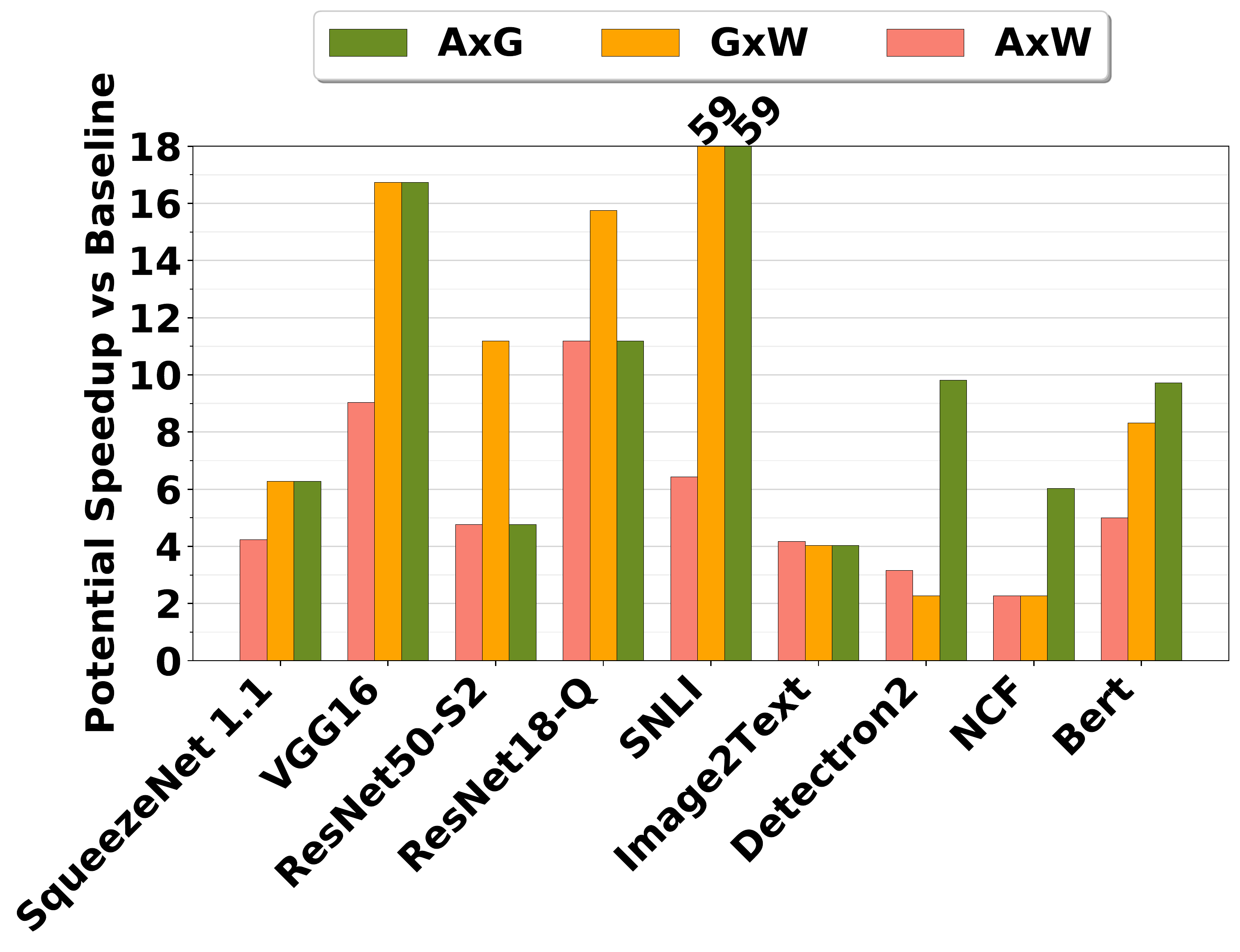}
  \vspace*{-0.5cm}
  \caption{Performance improvement potential of exploiting term sparsity for the three training phases.}
  \label{fig:potentials}
\end{minipage}
\vspace*{-0.5cm}
\end{figure*}

\vspace*{-0.2cm}

\section{Ineffectual Work In Training} \label{bit_sparsity}
This section motivates \OURL by measuring the work reduction that is theoretically possible with two related approaches: 1)~by removing all MACs where at least one of the operands are zero (value sparsity, or simply sparsity), and 2)~by processing only the non-zero bits of the mantissa of one of the operands (bit sparsity). 
We study the performance of \OURS on a wide variety of applications. Table~\ref{tbl:models} lists the models studied. {~ResNet18-Q is a variant of ResNet18 trained using PACT~\cite{PACT} which quantizes both activations and weights down to 4b during training.  ResNet50-S2 is a variant of ResNet50 trained using dynamic sparse reparameterization~\cite{dynSparse} which targets sparse learning that maintain high weight sparsity throughout the training process while achieving accuracy levels comparable to baseline training. SNLI performs natural language inference and comprises of fully-connected, LSTM-encoder, ReLU, and dropout layers. Image2Text is an encoder-decoder model for image-to-markup generation. We study three models of different tasks from MLPerf training benchmark: 1)~Detectron2: an object detection model based on Mask R-CNN, 2)~NCF: a model for collaborative filtering, and 3)~Bert: a transformer-based model using attention. For our measurements we sample one randomly selected batch per epoch over as many epochs as necessary to train the network to its originally reported accuracy (up to 90 epochs were enough for all).}

\begin{table}[]
\centering
\small
\caption{Models Studied}
\label{tbl:models}
\vspace*{-0.3cm}
\resizebox{\linewidth}{!}{%
\begin{tabular}{|l|l|l|}
\hline
\textbf{Model} & \textbf{Application}      & \textbf{Dataset}         \\ \hline
SqueezeNet 1.1 & Image Classification            & ImageNet~\cite{Imagenet} \\ \hline
VGG16          & Image Classification            & ImageNet~\cite{Imagenet}  \\ \hline
ResNet18-Q       &  Image Classification            & ImageNet~\cite{Imagenet} \\ \hline
ResNet50-S2   &  Image Classification            & ImageNet~\cite{Imagenet} \\ \hline
SNLI     & Natural Language Infer. & SNLI Corpus~\cite{snli} \\ \hline 
Image2Text  &   Image-to-Text Conversion &  im2latex-100k~\cite{im2latex-100k} \\ \hline
Detectron2  &   Object Detection & COCO~\cite{COCO} \\ \hline
NCF  &   Recommendation & ml-20m~\cite{ml-20m} \\ \hline
Bert  &   Language Translation &  WMT17~\cite{WMT17}\\ \hline
\end{tabular}}
\vspace{-0.5cm}
\end{table}

The bulk of work during training is due to three major operations per layer:

\noindent\begin{minipage}{.25\linewidth}
\boldmath
\begin{equation}\label{forward_eq}
    Z{=}I{\cdot}W
\end{equation}
\end{minipage}%
\begin{minipage}{.39\linewidth}
\boldmath
\begin{equation}\label{backward_eq1}
    \frac{\partial E}{\partial I}{=}W^{T}{\cdot}\frac{\partial E}{\partial Z}
\end{equation}
\end{minipage}%
\begin{minipage}{.36\linewidth}
\boldmath
\begin{equation}\label{backward_eq2}
    \frac{\partial E}{\partial W}{=}I{\cdot}\frac{\partial E}{\partial Z}
\end{equation}
\end{minipage}
\vspace{-0.3cm}
\break

For convolutional layers Eq.~\ref{forward_eq} describes the convolution of activations ($I$) and weights ($W$) that produces the output activations ($Z$) during forward propagation. There the output $Z$ passes through an activation function before used as input for the next layer. Eq.~\ref{backward_eq1} and~\ref{backward_eq2} describe the calculation of the activation ($\frac{\partial E}{\partial I}$) and weight ($\frac{\partial E}{\partial W}$) gradients respectively in the backward propagation. Only the activation gradients are back-propagated across layers. The weight gradients update the layer's weights once per batch. For fully-connected layers the equations describe several matrix-vector operations. For other operations they describe vector operations or matrix-vector operations. For clarity, in the rest of this work we will refer to the gradients as $G$.

Fig.~\ref{fig:value_sparsity}, and~\ref{fig:term_sparsity} show the value, and term-sparsity respectively and for each of the three tensors ($W$, $I$, and $G$). Each value is weighted according to frequency of use.  We use the term \textit{term-sparsity} to signify that for these measurements the mantissa is first encoded into signed powers of two using Canonical-encoding which is a variation of Booth-encoding. This is because of the bit-serial processing of the mantissa.

The activations in the image classification networks exhibit sparsity exceeding 35\% in all cases. This is expected since these networks use the ReLU activation function which clips negative values to zero. However, Weight sparsity is typically low and only some of the classification models exhibit sparsity in their gradients. For the remaining models, however, such as those for natural language processing, value sparsity is very low for all three tensors. Regardless, since some models do exhibit sparsity it may be worthwhile to investigate whether it possible to exploit it during training. As explained in the introduction, this is a non-trivial task, as training is different than inference and exhibits dynamic sparsity patterns on all tensors and different computation structure during the backward pass.

Fig.~\ref{fig:term_sparsity} shows that all three tensors exhibit high term-sparsity  for all models regardless of the target application. Given that term-sparsity is more prevalent than value sparsity, and exists in all models in the rest of this work we investigate whether it is practically possible to exploit it during training. One such option is the \OURL processing element which we present next. 

Fig.~\ref{fig:potentials} reports the ideal potential speedup due to reduction in the multiplication work through skipping the zero terms in the serial input. We calculate the potential speedup over the baseline as:
\vspace*{-0.2cm}
\begin{equation}
\resizebox{.9\hsize}{!}{$Potential\ speedup = \dfrac{\#MAC\ operations}{term\ sparsity \times \#MAC\ operations}$}
\end{equation}

\section{\OURLCORE\ Approach}\label{sec:design}
\OURL's goal is to take advantage of bit sparsity in one of the operands used in the three operations performed during training (equations~\ref{forward_eq} through~\ref{backward_eq2}) all of which are composed of many MAC operations.  We first explain how decomposing MAC operations into a series of shift-and-add operations can expose ineffectual work, providing us with the opportunity to save energy and time. Section~\ref{sec:PE} then describes the \OURL processing element and Section~\ref{sec:tile} details the \OURL tile.
\vspace*{-0.2cm}

\subsection{Exposing Ineffectual Work}\label{FP-PRA}
To expose ineffectual work during MAC operations we can decompose the operation into a series of ``shift and add'' operations, Let us first look at the multiplication. Let $A=2^{A_e}\times A_m$  and $B=2^{B_e} \times B_m$ be two values in floating point, both represented as an exponent ($A_e$ and $B_e$) and a significant ($A_m$ and $B_m$) which is normalized and includes the implied ``$1.$''. Conventional floating point units perform this multiplication in a single step (sign bits are XORed):
\begin{equation}
A \times B = 2^{A_e+B_e} \times (A_m \times B_m )= (A_m \times B_m) \ll (A_e +\ B_e) 
\end{equation}

By decomposing $A_m$ into a series $p$ of signed powers of two $A_m^p$ where $A=\sum_p A_m^p$ and $A_m^p = \pm 2^i$, we can instead perform the multiplication as follows: \\ \vspace*{-0.6cm}
\begin{equation}
\resizebox{.9\hsize}{!}{$A \times B = ( \sum_p A_m^p \times B_m ) \ll (A_e +\ B_e) 
=  \sum_p {\pm}B_m \ll (i\ + A_e\ +B_e) _{}$} 
\end{equation}

\begin{figure*}[h!]
\vspace{-1cm}
\begin{minipage}{.5\textwidth}
    \hspace{-1.5cm}
    \includegraphics[scale=0.8]{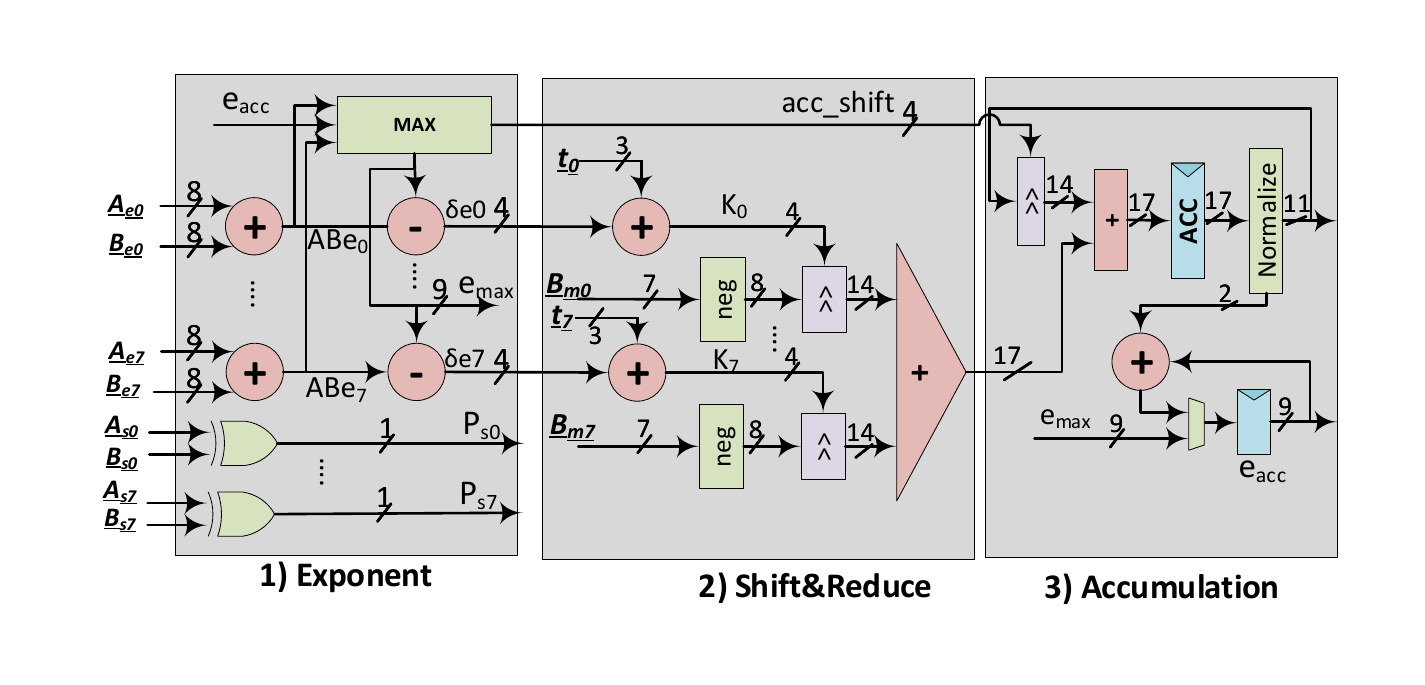}
    \vspace{-1cm}
    \caption{\OURL PE: Baseline Design.}
    \label{fig:FPRaker_PE}
\end{minipage}
\begin{minipage}{.45\textwidth}
    \hspace{-0.4cm}
    \includegraphics[scale=1.05]{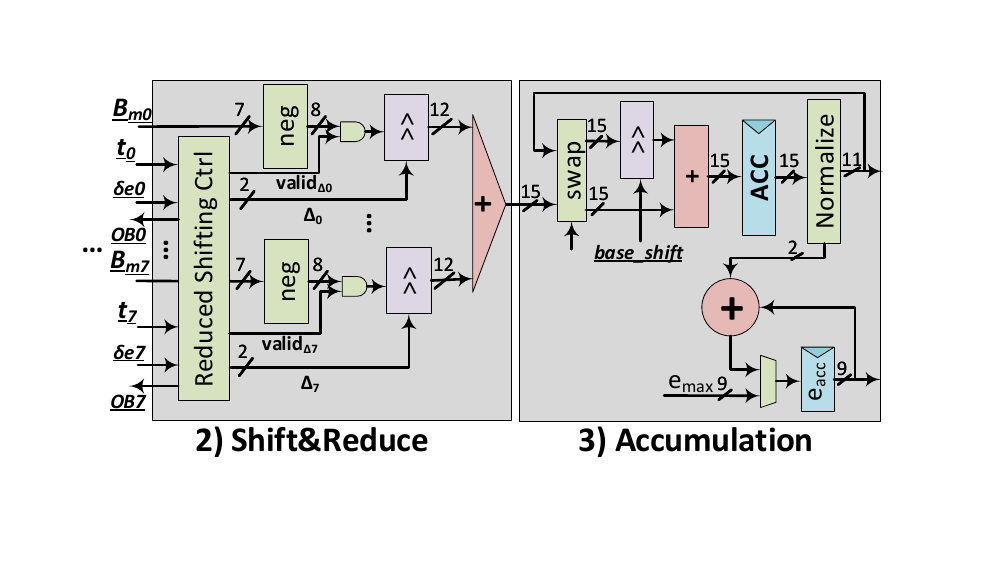}
    \vspace{-1.5cm}
    \caption{\OURL PE: Modified Design.}
    \label{fig:FPRaker_PE2}
\end{minipage}
\vspace{-0.5cm}
\end{figure*}
\vspace*{-0.4cm}

For example, if $A_m{=}1.0000001b$, $A_e{=}10b$, $B_m{=}1.1010011b$ and $B_e{=}11b$ then we can perform $A\times B$ as two shift-and-add operations of $B_m{\ll}(10b{+}11b{-}0)$ and $B_m{\ll}(10b{+}11b{-}111b)$. A conventional multiplier would process all bits of $A_m$ despite performing ineffectual work for the six bits that are zero. 

However, this decomposition exposes further ineffectual work that conventional units perform as a result of the high dynamic range of values that floating point seeks to represent. Informally, some of the work done during the multiplication will result in values that will be out-of-bounds given the accumulator value. To understand why this is the case let us now consider not only the multiplication but also the accumulation. Let's assume that the product $A\times B$ will be accumulated into a running sum $S$ and $S_e$ is much larger than $A_e + B_e$. It will not possible to represent the sum $S + A\times B$ given the limited precision of the mantissa. In other cases, some of the ``shift-and-add'' operations would be guaranteed to fall outside the mantissa even when we consider the increased mantissa length used to perform rounding, i.e., partial swamping. 
A conventional pipelined MAC unit can at best power-gate the multiplier and accumulator after comparing the exponents and only when the \textit{whole} multiplication result falls out of range. However, it cannot use this opportunity to reduce cycle count. By decomposing the multiplication into several simpler operations, we can terminate the operation in a single cycle given that we process the bits from the most to the least significand, and thus boost performance by initiating another MAC earlier. The same is true when processing multiple $A\times B$ products in parallel  in an adder-tree processing element. A conventional adder-tree based MAC unit can potentially power-gate the multiplier and the adder tree branches corresponding to products that will be out-of-bounds. The cycle will still be consumed. As we explain in the next section, a shift-and-add based unit will be able to terminate such products in a single cycle and advance others in their place.  

\section{\OURLCORE\ Implementation}\label{sec:design}
Section~\ref{sec:PE} describes the design of an \OURL processing element. {Section~\ref{sec:PE2} explains how \OURL time-multiplexes a single exponent block among multiple PEs, a key optimization for area- and energy-efficiency.} Section~\ref{sec:tile} and Section~\ref{sec:dataflow} explain respectively how multiple processing elements can be organized into a larger tile, and how data is organized in memory so that it can be supplied to the processing elements to keep them busy.

\subsection{\OURLCORE\ Processing Element}\label{sec:PE}

The \OURL PE performs the multiplication of 8 Bfloat16 $(A,B)$ value pairs, concurrently accumulating the result into an output accumulator. The Bfloat16 format consists of a sign bit, followed by a biased 8b exponent, and a normalized 7b significand (mantissa). Fig.~\ref{fig:FPRaker_PE} shows a baseline of the \OURL PE design which performs the computation in 3 blocks: exponent, reduction, and accumulation. We describe an implementation where the 3 blocks are performed in a single cycle. We will build upon this design and modify it to construct a more area efficient tile comprising several of these PEs. Recall that the significands of each of the $A$ operands are converted on-the-fly into a series of \textit{terms} (signed powers of two) using canonical encoding, e.g. $A{=}(1.1110000)$ is encoded as $(+2^{+1},-2^{-4})$. This encoding occurs just before the input to the PE. All values stay in bfloat16 while in memory. The PE will process the $A$ values term-serially.  The accumulator has an extended 13b significand; 1b for the leading 1 (hidden),  9b for extended precision following the chunk-based accumulation scheme as suggested by Sakr et al.,~\cite{IBMICLR2019} with a chunk-size of 64, plus 3b for rounding to nearest even. It has 3 additional integer bits following the hidden bit so that it can fit the worst case carry out from accumulating 8 products. In total the accumulator has 16b, 4 integer, and 12 fractional. 

The PE accepts 8 8-bit $A$ exponents $A_{e0},...,A_{e7}$, their corresponding 8 3-bit significand \textit{terms} $t_0,...,t_{7}$ (after canonical encoding) and signs bits $A_{s0},...,A_{s7}$, along with 8 8-bit $B$ exponents $B_{e0},...,B_{e7}$, their significands $B_{m0},...,B_{m7}$ (as-is) and their sign bits $B_{s0},...,B_{s7}$ as shown in Fig.~\ref{fig:FPRaker_PE}.

\textbf{Block 1 --- Exponent:}
Processing a new set of 8 value pairs starts first at the exponent block. This block adds the $A$ and $B$ exponents in pairs to produce the exponents $ABe_{i}$ for the corresponding products.  A comparator tree ($MAX$) takes these product exponents and the exponent of the accumulator and calculates the maximum $e_{max}$. The maximum exponent is used to align all products so that they can be summed correctly. To determine the proper alignment per product the block subtracts all product exponents from $e_{max}$ calculating the alignment offsets $\delta{e}_i$. The maximum exponent is used to also discard terms that will fall out-of-bounds when accumulated. The PE will skip any terms who fall outside the $e_{max}{-}12$ range. The block is invoked only \textit{once} per new set of value pairs,  \text{before} any terms are generated, and regardless of how many terms end up being generated. Accordingly, the minimum effective number of cycles for processing the 8 MACs will be 1 cycle regardless of value (the blocks can be pipelined, and since there are no data dependencies, the pipeline can be kept full). In case one of the resulting products has an exponent larger than the current accumulator exponent, the accumulator will be shifted accordingly prior to accumulation (\textit{acc\_shift} signal).

\begin{figure*}[h!]
    \vspace*{-0.5cm}
    \hspace*{-0.9cm}
    \includegraphics[scale=0.97]{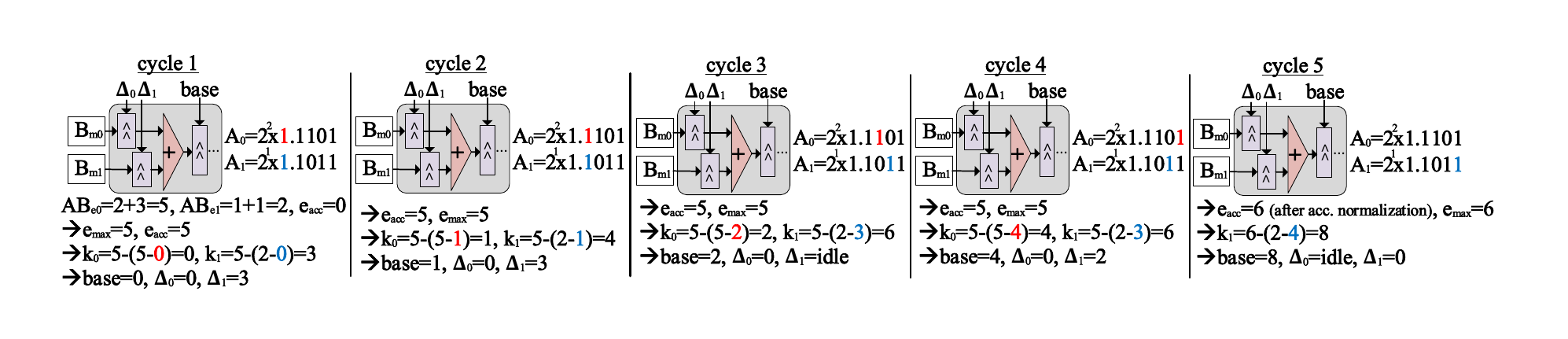} 
    \vspace{-1.2cm}
    \caption{\small{\OURL example: On cycle 1, the exponent block computes the products' exponents $AB_{e0}$ and $AB_{e1}$ (used once per new set of input pairs). Assuming a zero accumulator for simplicity, the MAX block computes $e_{max}{=}max(AB_{e0},AB_{e1},e_{acc}){=}5$. The accumulator exponent is then set to $e_{acc}{=}e_{max}$. The absolute terms are computed as $k_i{=}e_{acc}{-}(AB_{ei}{-}t_i)$, where $t_i$ are the position of non-zero terms generated by the term encoders shared across the PEs of the same tile column. To limit the PE's shifters range up to 3b, the shared shifter after the adder tree is set to $base{=}min(k_0,k_1)$. The value of the limited shifters are set to $\Delta_i{=}k_i{-}base$. For cycles 1 \& 2, since both $\Delta_0$ and $\Delta_1$ are within 3, both lanes can operate simultaneously. On cycle 3, the difference between $k_0$ and $k_1$ is more than 3, which means both terms cannot be processed simultaneously. Hence, lane 1 stalls while lane 0 operates with $base{=}k_0$ and $\Delta_0{=}0$. On cycle 4, lane 1 has its term from the previous cycle while lane 0 has a new term. Since the difference between the two terms is within 3, both lanes can operate simultaneously. On cycle 5, lane 0 is idle since it finished its terms while lane 1 processes its final term. To illustrate skipping out-of-bound terms, assume the total precision of the accumulator mantissa is 6b. On cycle 4, lane 1 feeds back a signal to term encoder indicating that any subsequent term coming from the same input pair is guaranteed to be ineffectual (out-of-bound) term. Hence, lane 1 can skip processing its last term and the PE saves one processing cycle by finishing at cycle 4.}}
    \label{fig:example}
    \vspace{-0.5cm}
\end{figure*}

\textbf{Block 2 --- Shift\&Reduce:} 
Since multiplication with a term amounts to shifting, this block calculates the number of bits by which each $B$ significand will have to be shifted by prior to accumulation. These are the 4-bit terms $K_0,...,K_{7}$. To calculate $K_i$ we add the product exponent deltas ($\delta{e}_i$) to the corresponding $A$ term $t_i$. The $A$ sign bits are XORed with their corresponding $B$ sign bits to determine the signs of the products $P_{s0},...,P_{s7}$.
The $B$ significands are complemented according to their corresponding product signs, and then shifted using the offsets $K_0,...,K_{7}$. The PE uses a shifter per $B$ significand to implement the multiplication.  In contrast, a conventional floating point unit would require shifters at the output of the multiplier. Thus \OURL PE effectively completely eliminates the cost of the multipliers.

Bits that are shifted out of the accumulator range from each $B_m$ operand are rounded using round-to-nearest-even (RNE) approach. An adder tree reduces the 8 $B_m$ operands into a single partial sum as shown in Fig.~\ref{fig:FPRaker_PE}(2).

\begin{figure}[ht]
    \vspace{-0.2cm}
    \centering
    \includegraphics[width=0.6\linewidth]{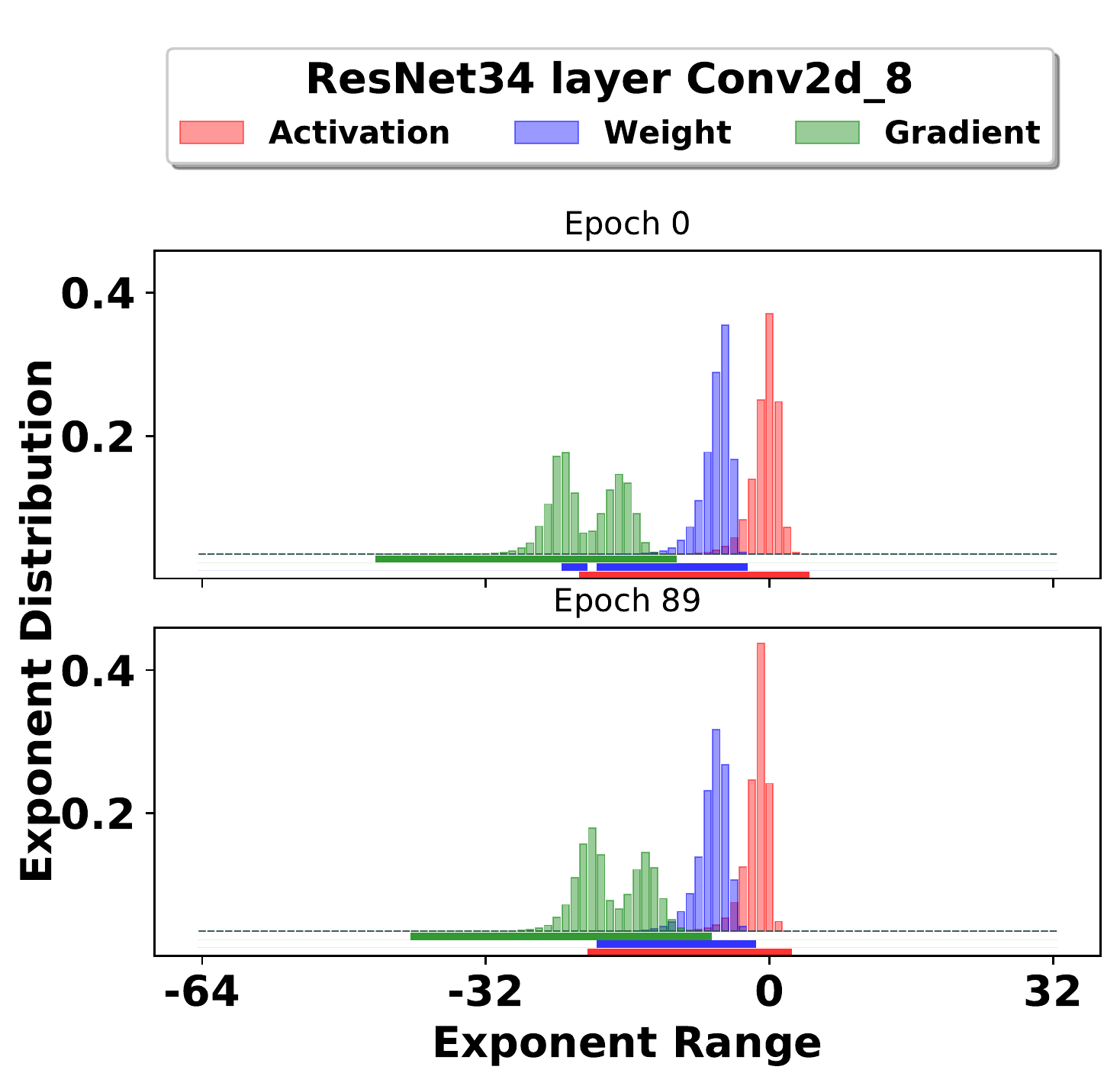} 
    \vspace{-0.3cm}
    \caption{Exponent histogram of layer $Conv2d\_8$ in epochs 0 and 89 of training ResNet34 on ImageNet. The figure shows only the utilized part of the full range [-127:128] of an 8b exponent.}
    \label{fig:exp_resnet34}
    \vspace{-0.3cm}
\end{figure}

\textbf{Block 3 --- Accumulation:} The resulting partial sum from step 2 is added to the correctly aligned value of the accumulator register. In each accumulation step, the accumulator register is normalized and rounded using the rounding-to-nearest-even (RNE) scheme. The normalization block updates the accumulator exponent as shown in Fig.~\ref{fig:FPRaker_PE}(3). When the accumulator value is read out, it is converted to bfloat16 by extracting only 7b for the significand.

In the worst case two $K_i$ offsets may differ by up to 12 since our accumulator has 12 fractional bits. This means that the baseline PE requires relatively large shifters and an accumulator tree that accepts wide inputs. Specifically, the PE requires shifters that can shift up to 12 positions a value that is 8b (7b significant + hidden bit). Had this been integer arithmetic we would need to accumulate 12+8=20b wide. However, since this is a floating point unit we will be accumulating only the 14 most significant bits (1b hidden, 12b fractional and the sign). Any bits falling below this range will be included in the sticky bit which is the least significant bit of each input operand. 

It is possible to further reduce this cost by taking advantage of the expected distribution of the exponents.  Fig.~\ref{fig:exp_resnet34} shows, for example, the distribution of exponents for a layer of ResNet34. The vast majority of the exponents of the inputs, the weights and the output gradients lie within a narrow range. This suggests that in the common case the exponent deltas will be relatively small. In addition, the MSBs of the activations are guaranteed to be one (given denormals are not supported~\cite{intel_bf16}). This indicates that very often the  $K_0,...,K_{7}$ offsets would lie within a narrow range. We take advantage of this behavior to reduce the PE area. In our preferred configuration we limit the maximum difference in among the $K_i$ offsets that can be handled in a \textit{single cycle} to be up to 3. As a result, the shifters need to support shifting by up to 3b and the adder tree now needs to process 12b inputs (1b hidden, 7b+3b significant, and the sign bit). A shared \textit{single} shifter ($base\_shift$) after the adder tree serves a dual purpose: First, it aligns the adder tree's output and the accumulator properly. Second, it allows the PE to skip over longer than 3b distances in the input term stream. This is useful, when the next set of terms are at a distance longer than 3b vs. the current ones. Fig.~\ref{fig:FPRaker_PE2} shows the modified PE. Each PE has a control unit to generate the modified terms $\Delta_i$ and a $valid_{\Delta i}$ signal indicating whether the lane can process its term at the current cycle.

Skipping out-of-bounds terms turns out to be surprisingly inexpensive. The control unit uses a comparator per lane to check if its current $K$ term lies within a threshold with the value of the accumulator precision (comparators are optimized by the synthesis tool for comparing with a constant) and feeds back an $OB_i$ signal to its corresponding term encoder indicating that any subsequent term coming from the same input pair is guaranteed to be ineffectual (out-of-bound) given the current $e_{acc}$ value. Hence, \OURL can boost its performance and energy-efficiency by skipping the processing of the subsequent out-of-bound terms. The $OB_i$ signals of a certain lane across the PEs of the same tile column are synchronized together. The threshold is currently set according to~\cite{IBMICLR2019} which ensures models converge within 0.5\% of the FP32 training accuracy on ImageNet dataset. However, the threshold can be controlled effectively implementing a dynamic bit-width accumulator which can boost performance by increasing the number of skipped "out-of-bounds" bits, an option we do not investigate further. Fig.~\ref{fig:example} shows an example of a simplified PE processing 2 activation-weight pairs each with 4b mantissa: $A_0{=}2^2{\times}1.1101, B_0{=}2^3{\times}1.0011$ and $A_1{=}2^1{\times}1.1011, B_1{=}2^1{\times}1.1010$.

\subsection{Sharing the Exponent Block} \label{sec:PE2}

\begin{figure}[ht]
    \vspace{-0.6cm}
    \includegraphics[width=0.95\linewidth]{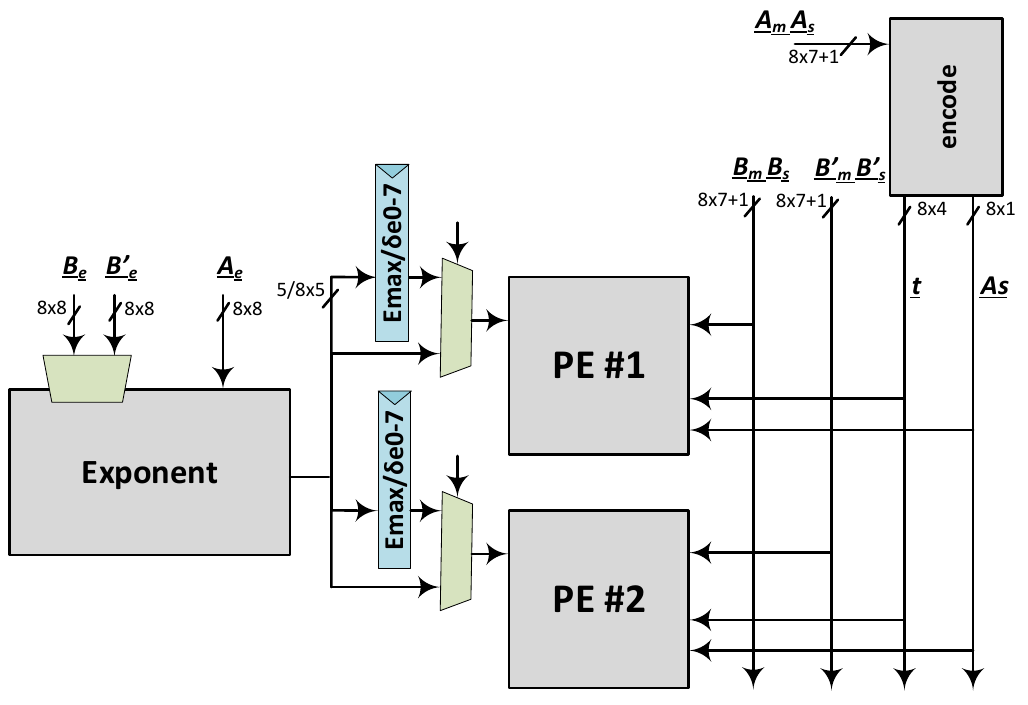} 
    \vspace{-0.3cm}
    \caption{Reducing area by sharing the exponent block between two PEs.}
    \label{fig:PE2}
    \vspace{-0.5cm}
\end{figure}

In the common case, processing a group of $A$ values will require multiple cycles since some of them will be converted into multiple terms. During that time the inputs to the exponent block will not change. To further reduce area we can take advantage of this expected behavior and share the exponent block across multiple PEs. {The decision of how many PEs to share an exponent block over can be based on the expected bit-sparsity. The lower the bit-sparsity then higher the processing time per PE and the less often it will need a new set of exponents. Hence, the more the PEs that can share the exponent block. Since the studied models are highly sparse, sharing one exponent block per two PEs proved best.} Figure~\ref{fig:PE2} shows the modified design. The unit as a whole accepts as input one set of 8 $A$ inputs and two sets of $B$ inputs, $B$ and $B'$. The exponent block can process one of $(A,B)$ or $(A,B')$ at a time. During the cycle when it processes $(A,B)$ the multiplexer for PE\#1 passes on the $e_{max}$ and exponent deltas directly to the PE. Simultaneously, these values will be latched into the registers in front of the PE so that they remain constant while the PE processes all terms of input $A$. When the exponent block processes $(A,B')$ the aforementioned process proceeds with PE\#2. With this arrangement both PEs must finish processing all $A$ terms before they can proceed to process another set of $A$ values. Since the exponent block is shared, each set of 8 $A$ values will take at least 2 cycles to be processed (even if it contains zero terms).

\subsection{FPRaker Tile} \label{sec:tile}

\begin{figure}[H]
    \vspace*{-0.9cm}
    \includegraphics[width=\linewidth]{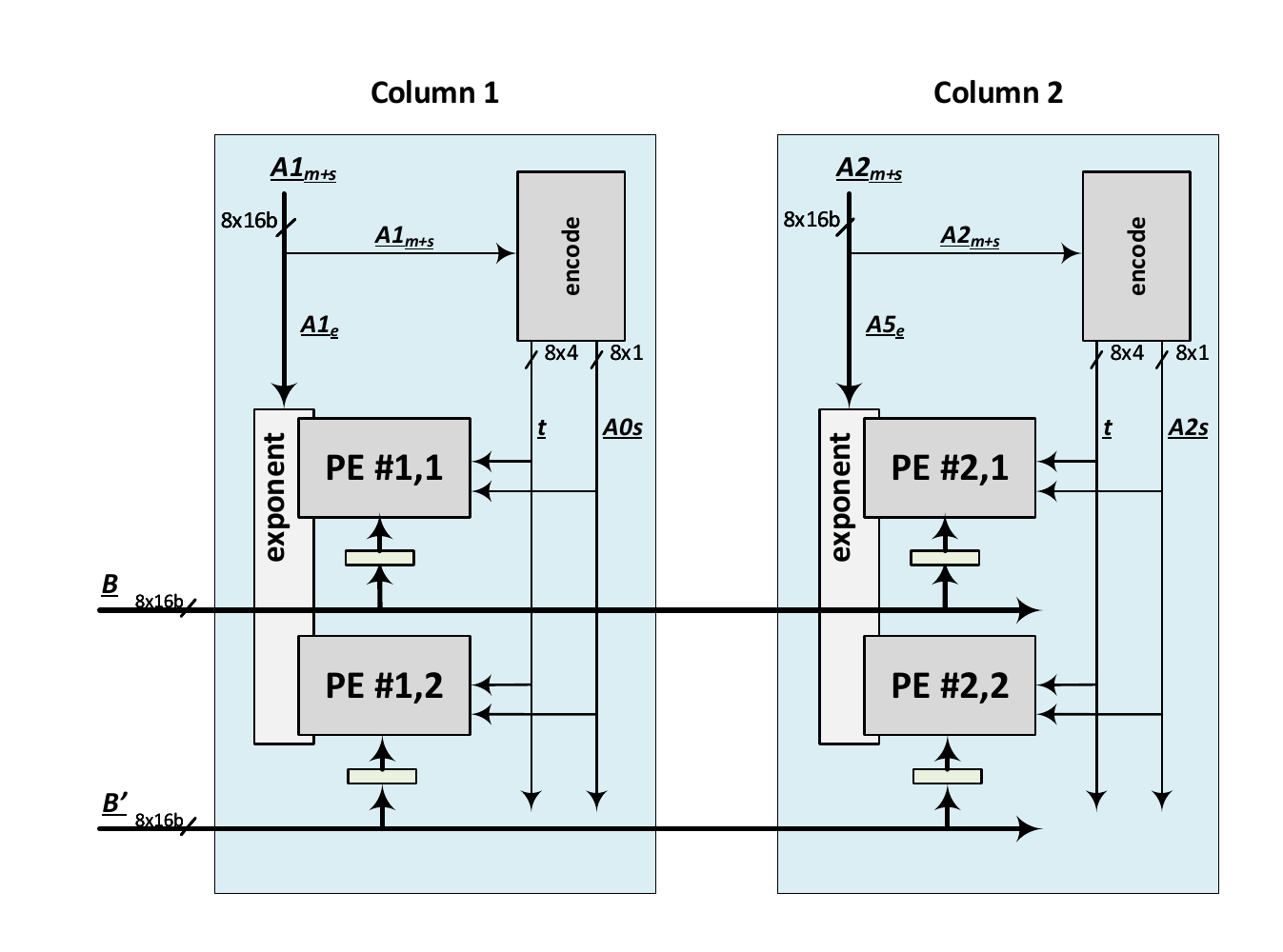} 
    \vspace{-0.8cm}
    \caption{A $2\times2$ PE \OURL Tile.}
    \label{fig:tile}
    \vspace{-0.3cm}
\end{figure}

By utilizing per PE buffers it is possible to exploit data reuse temporally. To exploit data reuse spatially we can arrange several PEs into a tile. Fig.~\ref{fig:tile} shows an example of a $2\times2$ tile of PEs and each PE performs 8 MAC operations in parallel. Each pair of PEs per column shares an exponent block as previously described. The $B$ and $B'$ inputs are shared across PEs in the same row. For example, during the forward pass we can have different filters being processed by each row and different windows processed across the columns.  Since the $B$ and $B'$ inputs are shared all columns would have to wait for the column with the most $A_i$ terms to finish before advancing to the next set of $B$ and $B'$ inputs. To reduce these stalls the tile introduces per $B$ and $B'$ buffers. By having $N$ such buffers per PE allows the columns be at most $N$ sets of values ahead. 

\subsection{Exponent Base-Delta Compression}\label{sec:compression}
\begin{figure}[H]
    \centering
    \vspace{-0.8cm}
    \includegraphics[scale=0.8]{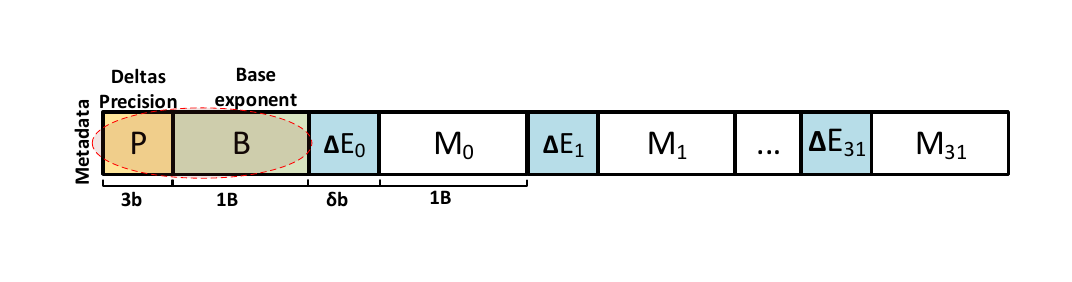} 
    \vspace{-0.5cm}
    \caption{Group of 32 values with exponent base-delta compression}
    \label{fig:BDcompression_group}
    \vspace{-0.4cm}
\end{figure}

{Motivated by the narrow value distribution shown in Fig. \ref{fig:exp_resnet34}, we studied the spatial correlation of values during training. We found that consecutive values across all dimenions (channel, H, or W) have similar values. This is true for the activations, the weights and the output gradients. Similar values in floating-point have similar exponents, a property which we can exploit through a base-delta compression scheme~\cite{BDcompression}. In our experiments, we block values channel-wise (we present evidence, however, that the value corellation persists along the H dimension as well) into groups of 32 values each, where the exponent of the first value in the group is the base and we compute the delta exponent for the rest of the values in the group relative to it as shown in Fig.~\ref{fig:BDcompression_group}. The bit-width ($\delta$) of the delta exponents is dynamically determined per group and is set to the maximum precision (P) of the resulting delta exponents per group similar to the approach of Delmas et al.~\cite{Shapeshifter}. The delta exponent bit-width (3b) is attached to the header of each group as metadata.}

\begin{figure}[H]
    \vspace*{-0.3cm}
    \centering
    \includegraphics[width=0.8\linewidth]{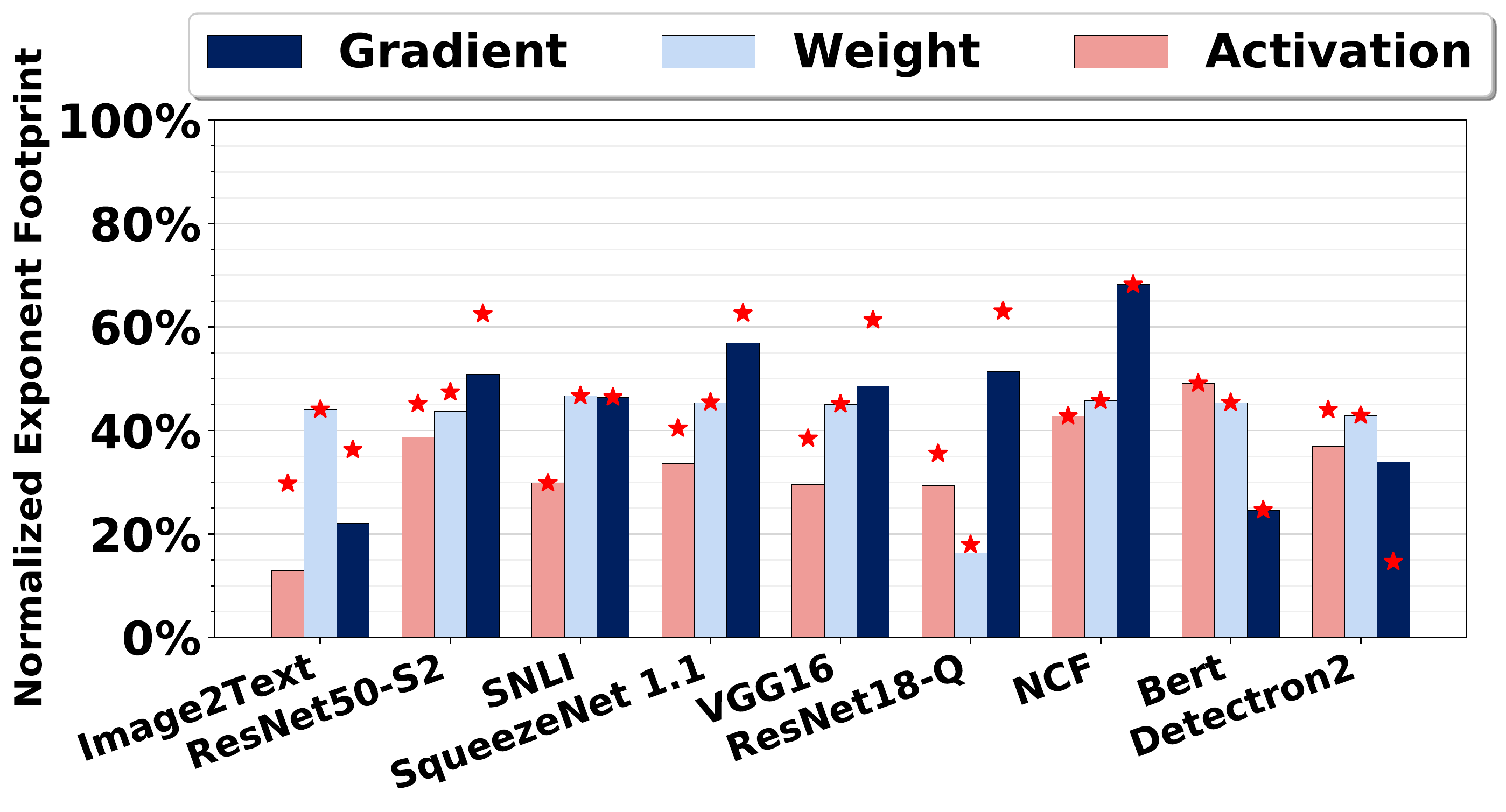} 
    \vspace{-0.3cm}
    \caption{\hl{Memory savings due to exponent base-delta compression. Bars and markers represent compression channel-wise and spatial-wise, respectively.}}
    \label{fig:compression}
    \vspace*{-0.3cm}
\end{figure}

{Fig.~\ref{fig:compression} shows the total, normalized exponent footprint after the base-delta compression. Our technique is effective for both channel-wise and spatial (H dimension shown) dataflows. We use this compression scheme to reduce the off-chip memory bandwidth. Values are compressed at the output of each layer and before writing them off-chip, and they are decompressed when they are read back on-chip.}

\subsection{Data Supply}\label{sec:dataflow}
Focusing solely on computation is insufficient. Data transfers account for a significant portion and often dominate energy consumption in deep learning. Accordingly, it is essential to consider what the memory hierarchy needs to do to keep the execution units busy. A challenge with training is that while it processes three arrays $I$, $W$ and $G$ the order in which their elements are grouped differs across the three major computations (Eq.~\ref{forward_eq} through~\ref{backward_eq2}). However it is possible to rearrange the arrays as they are read from off-chip. For this purpose we store the arrays in memory using a container of ``square'' of $32{\times}32$ bfloat16 values. This a size that matches well the typical row sizes of DDR4 memories and allows us to achieve high bandwidth when reading values from off-chip. A container includes values from coordinates $(c,r,k)$ (channel, row, column) to $(c+31,r,k+31)$ where $c$ and $k$ are divisible by 32 (padding is used as necessary). Containers are stored in channel, column, row order. When read from off-chip memory the container values are stored either in the exact same order on the multibanked on-chip buffers. The tiles can then access data directly reading 8 bfloat16 values per access. The weights and the activation gradients however need to be processed in different order depending on the operation performed. Effectively, the respective arrays must be accessed in the transpose order during one of the operations. For this purpose we incorporate transposer units on-chip. A transposer reads in 8 blocks of 8 bfloat16 values from the on-chip memories. Each of these 8 reads uses 8-value wide reads as mentioned above and the blocks are written as rows in an internal to the transposer buffer. Collectively these blocks form an $8\times 8$ block of values. The transposer can read out 8 blocks of 8 values each and send those to the processing units. Each of these blocks however is read out as a column from its internal buffer. This effectively transposes the $8\times 8$ value group.

\vspace*{-0.2cm}

\section{Evaluation}\label{sec:eval}
This section evaluates the performance of \OURS comparing against an equivalent baseline architecture that uses conventional floating-point units. The rest of this section is organized as follows: Section~\ref{sec:methodology} presents the experimental methodology. Section~\ref{sec:performance} evaluates the performance of \OURS over the baseline architecture. Section~\ref{sec:energy} reports the energy efficiency.
\subsection{Methodology}\label{sec:methodology}
A custom cycle-accurate simulator was developed to model the execution time of \OURS and of the baseline architecture. Besides modeling timing behavior the simulator also models value transfers and computation in time faithfully and checks the produced values for correctness against the golden values. The simulator was validated with microbenchmarking. For area and power analysis, both \OURS and the baseline designs were implemented in Verilog and synthesized using Synopsys' Design Compiler~\cite{synopsysDC} for 600 MHz clock frequency with a 65nm TSMC technology (due to licensing restrictions we cannot get access to a better technology) and with a commercial library for the given technology. We use Cadence Innovus \cite{Innovus} for layout generation. We use Intel's PSG ModelSim to generate data-driven activity factors which we feed to Innovus to estimate the power. The baseline MAC unit was optimized for area, energy and latency. Generally, it is not possible to optimize for all three, however, in the case of MAC units, it is possible~\cite{DBLP:journals/tc/GalalH11}. {We use an efficient bit-parallel fused MAC unit as the baseline PE. The constituent multipliers are both area and latency efficient, and are taken from the DesignWare IP library developed by Synopsys. Further, we optimize the baseline units for deep learning training by reducing the precision of its I/O operands to bfloat16 and accumulating in reduced precision with chunk-based accumulation similar to FPRaker units.}
The area and energy consumption of the on-chip SRAM Global Buffer (GB) is divided into activation, weight, and gradient memories which were modeled using CACTI ~\cite{cacti}. The Global Buffer has an odd number of banks to reduce bank conflicts for layers with a stride greater than one. To estimate the latency and energy consumption of the off-chip DRAM memory we use the model provided by Micron ~\cite{micron}. The configurations for both \OURS and the baseline are shown in Table~\ref{tbl:accelerator:configs}.

 \begin{table}
		\centering\small
		\caption{Baseline and \OURS configurations.}
		\vspace{-0.3cm}
		\label{tbl:accelerator:configs}
		\footnotesize
		 \begin{tabular}{|c|c|c|}
         \hline
         & \textbf{\OURLCORE} & \textbf{Baseline} \\ \hline
         \hline
         \textbf{Tile Configuration} & $8 \times 8$ & $8 \times 8$\\ \hline
         \textbf{Tiles} & 36 & 8 \\ \hline
         \textbf{Total PEs} & 2304 & 512 \\ \hline
         \textbf{Multipliers/PE} & 8  & 8 \textsc{bfloat16}\\ \hline
         \textbf{MACs/cycle} & - & 4096 \\ \hline
         \textbf{Scratchpads} & \multicolumn{2}{c|}{2KB each} \\ \hline
         \textbf{Global Buffer} & \multicolumn{2}{c|}{4MB $\times$ 9 banks} \\\hline
         \textbf{Off-chip DRAM Memory} & \multicolumn{2}{c|}{16GB 4-channel LPDDR4-3200} \\\hline
		\end{tabular}
		\vspace*{-0.3cm}
	\end{table}


To evaluate our accelerator, we collected traces for one random mini-batch during the forward and backward pass in each epoch of training. All models were trained long enough to attain the maximum top-1 accuracy as reported by the original authors. To collect the traces, we trained each model on a NVIDIA RTX 2080 Ti GPU and stored all of the inputs and outputs for each layer using Pytorch Forward and Backward hooks. For BERT we traced BERT-base and the fine-tuning training for a GLUE task. The simulator uses the traces to model execution time and collects activity statistics so that energy can be modeled.
\vspace*{-0.2cm}

\subsection{Area}\label{PRA_area}
Since \OURL processes one of the inputs term-serially a single \OURL processing engine can never outperform a conventional PE that processes the same number of inputs. \OURL relies on parallelism to extract more performance. This is only possible if we can afford to use more \OURL units than conventional units. One approach is to use an iso-compute area constraint. That is to determine how many \OURL tiles we can fit in the same area for a baseline tile. For this we take into account only the compute cores as associated logic and not the scratchpads. 

The conventional PE we compared against process concurrently 8 pairs of bfloat16 values and accumulates their sum. We can include buffers for the inputs (A and B) and the outputs so that we can exploit data reuse temporally. We can also organize multiple PEs in a grid sharing buffers and inputs across rows and columns to also exploit reuse spatially. {Both \OURL and the baseline are configured to have scaled-up GPU Tensor-Core-like tiles that perform $8\times8$ vector-matrix multiplication where 64 PEs are organized in a $8\times8$ grid and each PE performs 8 MAC operations in parallel.}

{Post layout, and taking into account only the compute area, an \OURL tile occupies 22\% the area vs. the baseline tile. Table~\ref{tbl:area_power} reports the corresponding area and power per tile. Accordingly, to perform an iso-compute-area comparison, we configure the baseline accelerator to have 8 tiles and \OURL with 36 tiles.} The area for the on-chip SRAM global buffer is $344 mm^2$, $ 93.6mm^2$, and $334 mm^2$ for the activations, weights, and gradients, respectively.


 \begin{table}
		\centering\small
		\caption{Breakdown of the area and power consumption per tile of \OURS vs. Baseline.}
		\label{tbl:area_power}
		\vspace{-0.3cm}
		\resizebox{\linewidth}{!}{%
		 \begin{tabular}{|c|c|c|c|c|}
         \hline
         \multicolumn{5}{|c|}{\textbf{\texttt{Area [\SI{}{\micro\meter^2}]}}} \\ \hline
         & \textbf{PE Array} & \textbf{Term Encoders} & \textbf{Total} & \textbf{Normalized} \\ \hline
         \textbf{FPRaker} & 304,118 & 12,950 & 317,068 & $0.22\times$ \\ \hline
         \textbf{Baseline} & 1,421,579 & N/A & 1,421,579 & $1\times$\\ \hline
         \multicolumn{5}{|c|}{\textbf{\texttt{Power [$mW$]}}} \\  \hline 
         \textbf{FPRaker} & 104 & 5.5 & 109.5 & $0.23\times$ \\  \hline 
         \textbf{Baseline} & 475 & N/A & 475 & $1\times$ \\  \hline
         \multicolumn{5}{|c|}{\textbf{\texttt{Energy Efficiency Normalized to Baseline}}} \\ \hline
         \multicolumn{3}{|c|}{\textbf{FPRaker}} & \multicolumn{2}{c|}{$1.75\times$} \\ \hline
		\end{tabular}}
		\vspace*{-0.5cm}
	\end{table}

\begin{figure*}[h!]
\begin{minipage}{.35\textwidth}
  \includegraphics[scale=0.18]{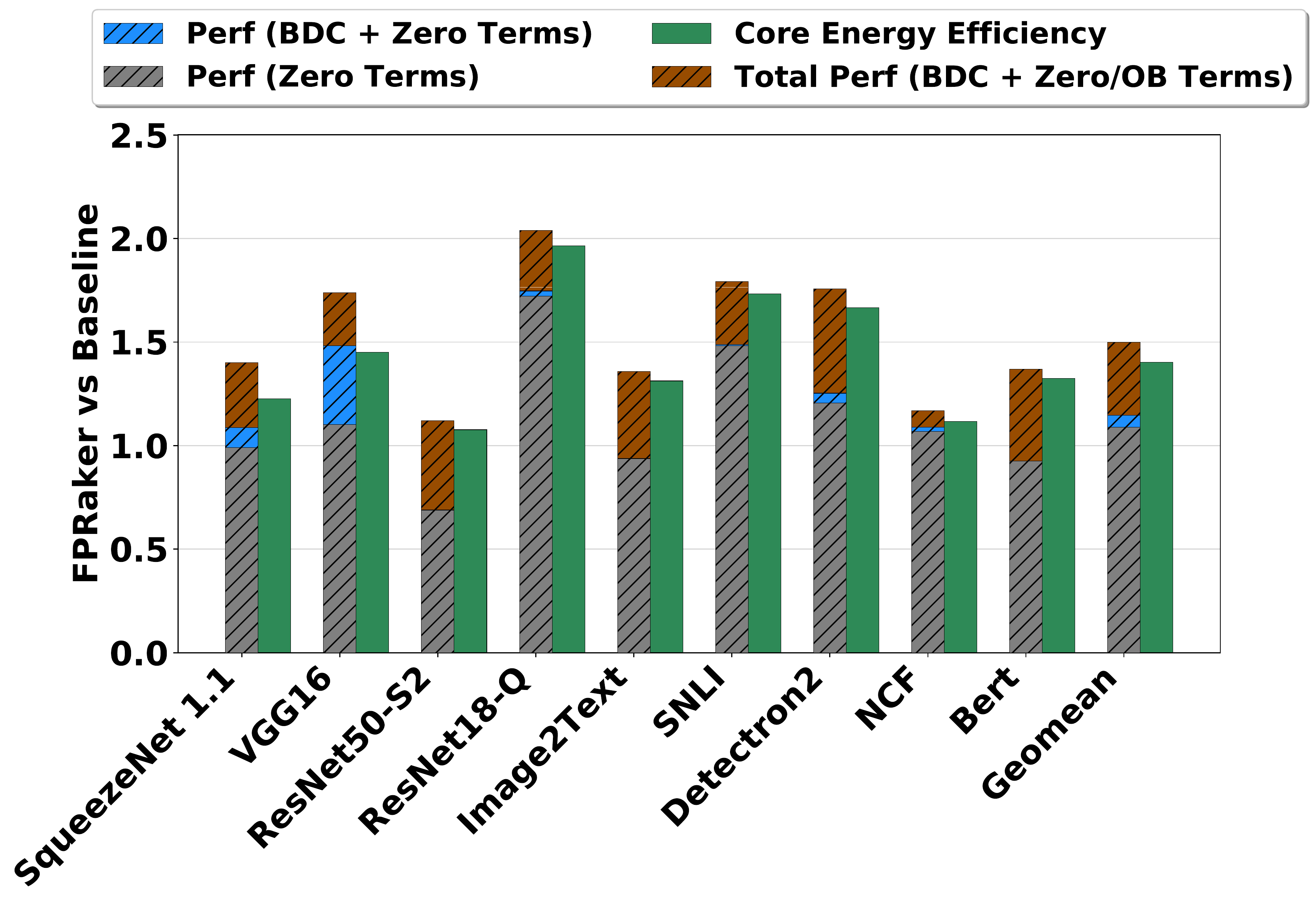}
  \vspace*{-0.2cm}
  \caption{iso-compute-area performance and energy-efficiency comparison between \OURS and the baseline.}
  \label{fig:total_speedup_energy}
\end{minipage}
\hspace*{0.1cm}
\begin{minipage}{.35\textwidth}
  \vspace*{-1cm}
  \includegraphics[scale=0.17]{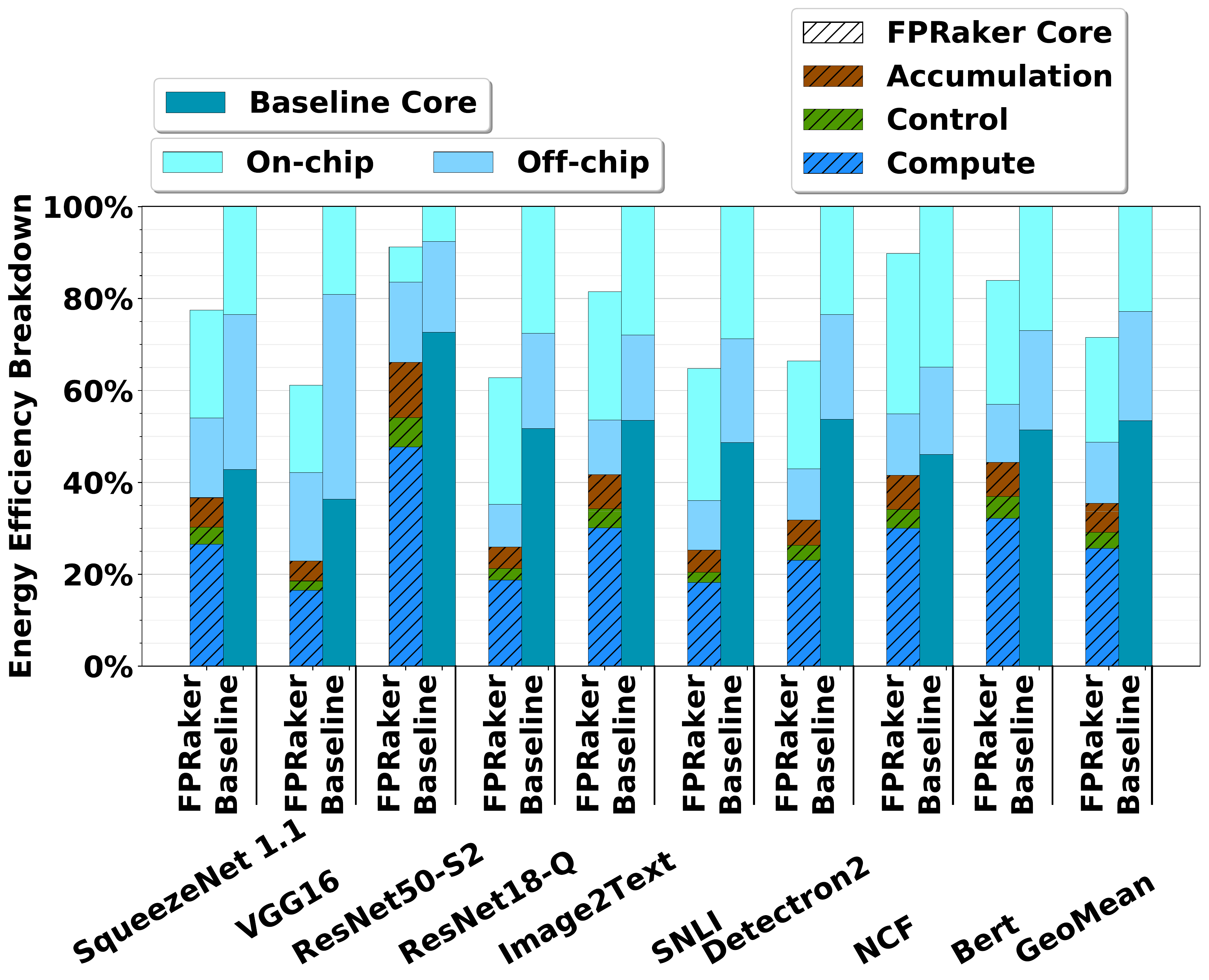}
  \vspace*{-0.2cm}
  \caption{Overall Energy Efficiency\\ of \OURS vs. Baseline.}
  \label{fig:energy_eff_breakdown}
\end{minipage}
\hspace*{-0.3cm}
\begin{minipage}{.26\textwidth}
  \includegraphics[scale=0.185]{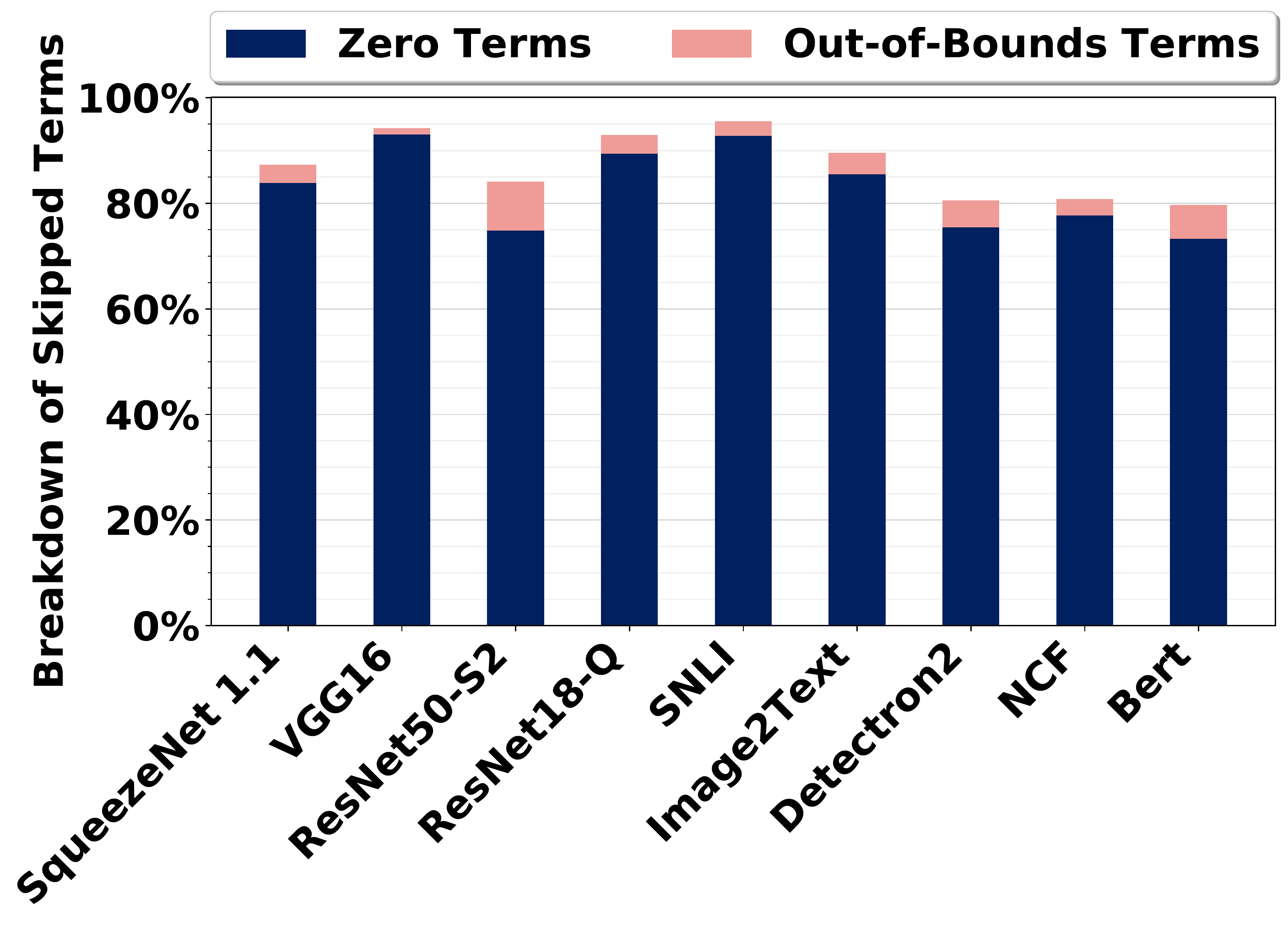}
  \caption{Breakdown of skipped terms by \OURS.}
  \label{fig:benefits_breakdown}
\end{minipage}
\vspace{-0.5cm}
\end{figure*}
\vspace*{-0.1cm}

\subsection{Execution Time}\label{sec:performance}
Figure~\ref{fig:total_speedup_energy} shows the performance breakdown of \OURL due to the base-delta compression (BDC), and skipping zero and out-of-bound (OB) terms relative to the baseline. On average, \OURS outperforms the baseline by $1.5\times$ (skipping zero terms: $+9\%$, BDC: $+5.8\%$, skipping out-of-bound terms: $+35.2\%$). From the studied convolution-based models, ResNet18-Q benefits the most from \OURS where the performance improves by  $2.04\times$ over the baseline. Training for this network incorporates PACT quantization and as a result most of the activations and weights throughout the training process can fit in 4b or less. This translates into high term sparsity which \OURS exploits. This result demonstrates that \OURL can deliver benefits with specialized quantization methods without requiring that the hardware be also specialized for this purpose. 

 SNLI, NCF, and Bert are dominated by fully connected layers. {while in fully-connected layers there is no weight reuse among different output activations, training can take advantage of batching to maximize weight reuse across multiple inputs (e.g., words) of the same input sentence which results in higher utilization of the tile PEs.  Speedups follow bit sparsity. For example, \OURL achieves a speedup of $1.8\times$ over the baseline for SNLI due its high bit sparsity.}


\vspace*{-0.2cm}

\subsection{Energy Efficiency}\label{sec:energy}
Figure \ref{fig:total_speedup_energy} shows the total energy efficiency of \OURS over the baseline architecture for each of the studied models. On average, \OURS is $1.4\times$ more energy efficient compared to the baseline considering only the core logic and $1.36\times$ more energy efficient when everything is taken into account. The energy-efficiency improvements follow closely the performance benefits. For example, benefits are higher at around $1.7\times$ for SNLI and Detectron2. The quantization in ResNet18-Q boosts the core logic energy efficiency to as high as $1.97\times$. Fig.~\ref{fig:energy_eff_breakdown} shows the energy consumed by \OURL normalized to the baseline as a breakdown across three main components: core logic, off-chip and on-chip data transfers. \hl{Fig.}~\ref{fig:energy_eff_breakdown} \hl{further breaks down} \OURL \hl{core into ``Compute'' (PE stages 1 and 2), ``Control'' (PE control units and shared term encoders), and ``Accumulation'' (PE stage 3).} \OURL along with the exponent base-delta compression reduce the energy consumption of the core logic and off-chip memory significantly.
\vspace*{-0.2cm}

\subsection{Performance Analysis}

\noindent{\textbf{Skipped Terms: }}Figure~\ref{fig:benefits_breakdown} shows a breakdown of the terms \OURL skips. There are two cases: 1)~skipping zero terms, and 2)~skipping non-zero terms that are out-of-bounds due to the limited precision of the floating-point representation. 

Skipping out-of-bounds terms increases term sparsity for ResNet50-S2 and Detectron2 by around 10\% and 5.1\%, respectively. Networks with high sparsity (zero values) such as VGG16 and SNLI benefit the least from skipping out-of-bounds terms with the majority of term sparsity coming from zero terms. This is because there are few terms to start with. For ResNet18-Q, most benefits come from skipping zero terms as the activations and weights are effectively quantized to 4b values. However, as Figure~\ref{fig:total_speedup_energy} \hl{showed, skipping out-of-bound terms improves performance much more than the fraction of terms skipped would suggest. Recall, that all lanes must wait for the slowest one to finish processing amplifying the effect on performance across all lanes. In the worse case, all other 7 lanes are waiting for a single one to finish, wasting 7 execute slow. Skipping out-of-bound terms reduces this synchronization overhead. }

\begin{figure*}[h!]
\begin{minipage}{.29\textwidth}
    \includegraphics[scale=0.175]{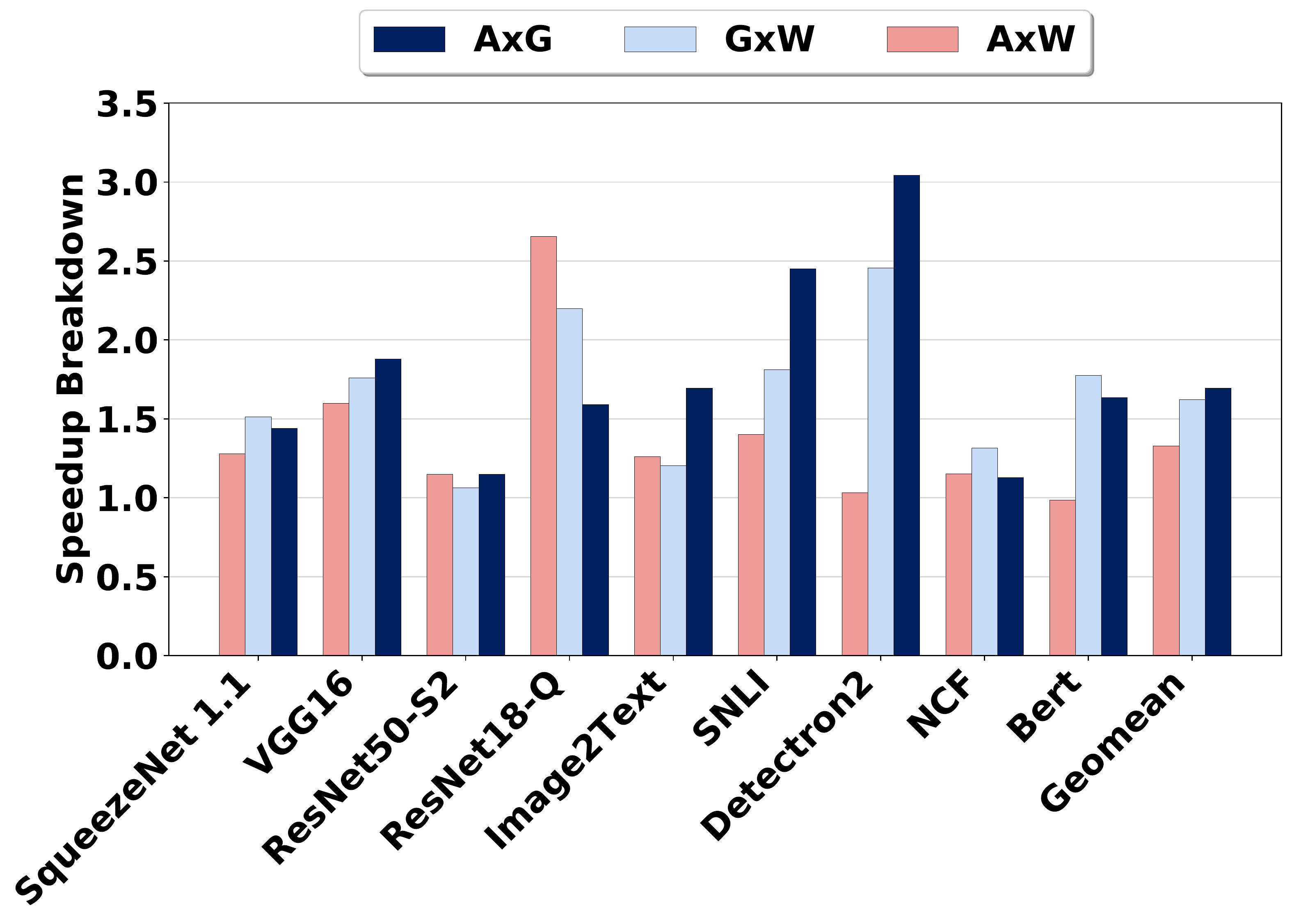}
    \vspace{-0.4cm}
    \caption{Breakdown of \OURS speedup over the baseline.}
    \label{fig:speedup_per_phase}
\end{minipage}
\hfill
\begin{minipage}{.33\textwidth}   
  \hspace{0.15cm}
  \includegraphics[scale=0.2]{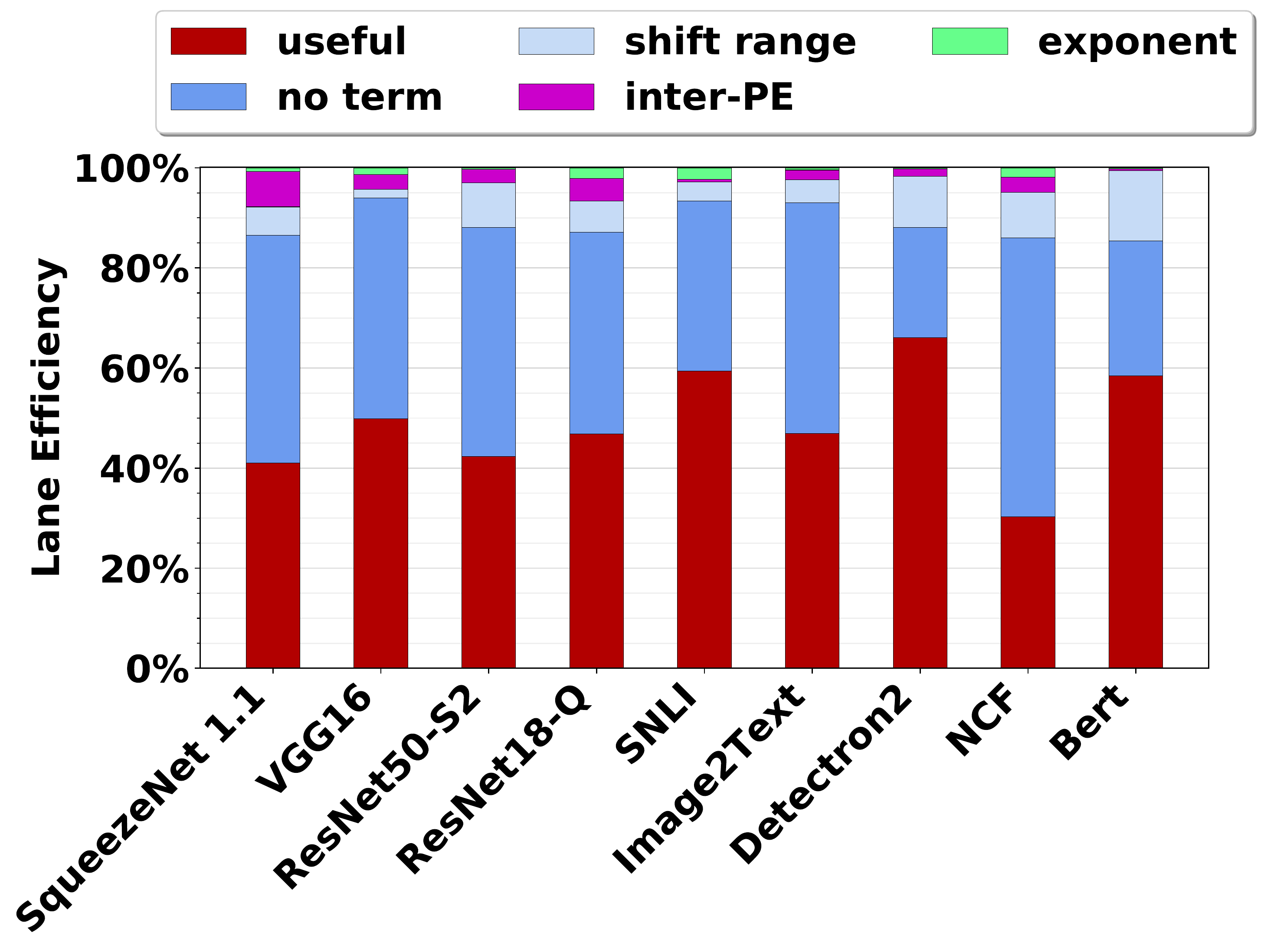}
  \vspace*{-0.7cm}
  \caption{Breakdown of execution cycles of \OURS.}
  \label{fig:cycles_breakdown}
\end{minipage}
\hfill
\begin{minipage}{.35\textwidth}  
  \hspace*{0.2cm}
  \includegraphics[scale=0.18]{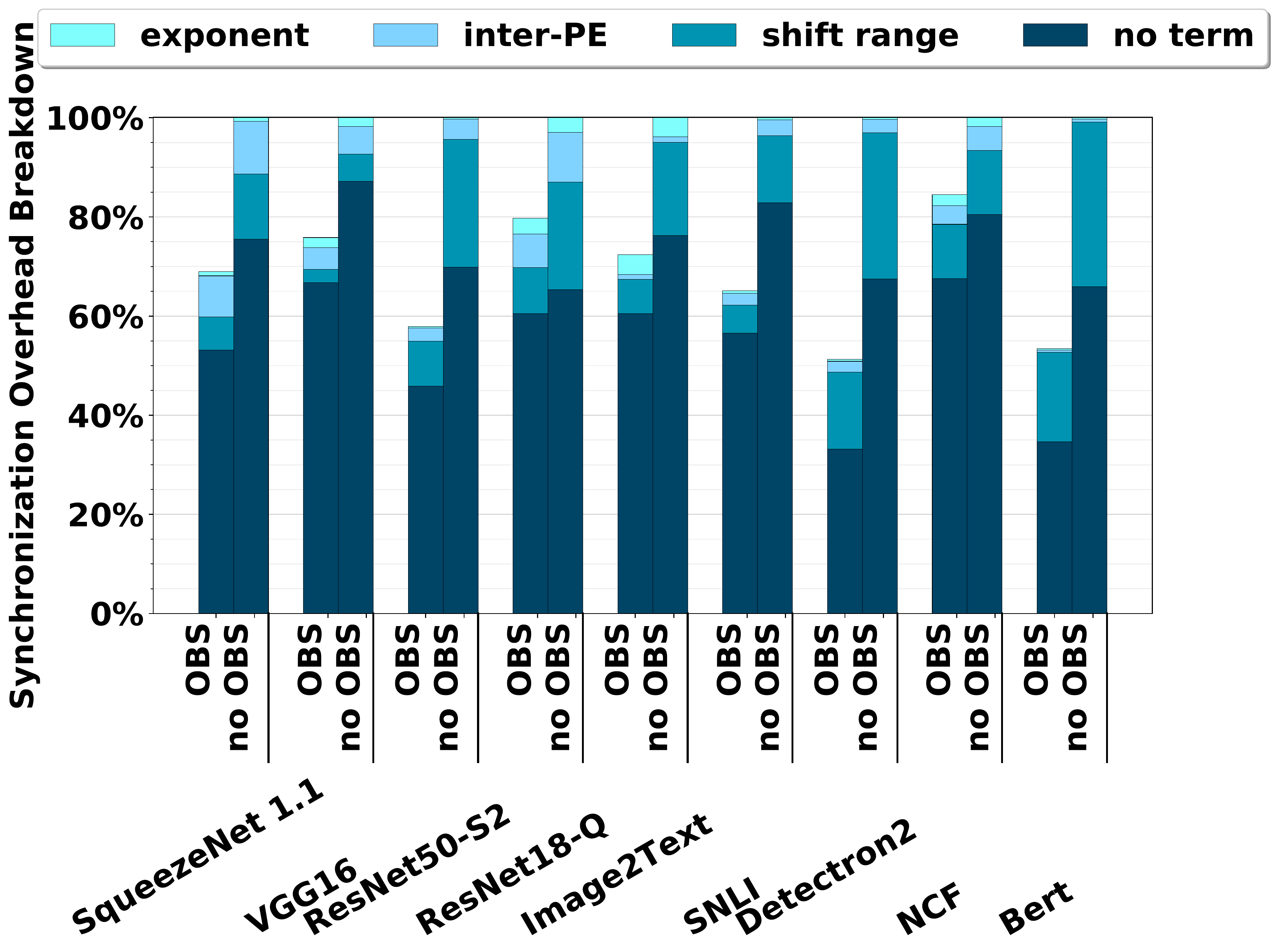}
  \vspace*{-0.7cm}
  \caption{\hl{Effect of out-of-bound terms skipping (OBS) on the synchronization overhead.}}
  \label{fig:sync_overhead_breakdown}
\end{minipage}
\vspace{-0.5cm}
\end{figure*}

\noindent{\textbf{Computation Phase: }}Figure~\ref{fig:speedup_per_phase} reports speedup for each of the 3 phases of training: the $A\times W$ in forward propagation, and the $A\times G$ and the $G\times W$ to calculate the weight and input gradients in the backpropagation, respectively. \OURS consistently outperforms the baseline for all three phases. The speedup depends on the amount of term sparsity, and the value distribution of $A$, $W$, and $G$ across models, layers, and training phases. The less terms a value has the higher the potential for \OURL to improve performance. However, due to the limited shifting that the \OURL PE can perform per cycle (up to 3 positions) how terms are distributed within a value impacts the number of cycles needed to process it. This behavior applies across lanes to the same PE and across PEs in the same tile. In general, the set of values that are processed concurrently will translate into a specific term sparsity pattern. \OURL favors patterns where the terms are close to each other numerically. 

\noindent{\textbf{Where Cycles Go: }} Figure \ref{fig:cycles_breakdown} reports a breakdown of PE lane utilization and highlights performance bottlenecks. There are several reasons why stalls occur: 1)~inter-PE synchronization, 2)~intra-PE synchronization, and 3)~Sharing the exponent block. Intra-PE synchronization stalls can happen in two scenarios: a)~imbalance in the number of terms assigned for each lane within a PE due to uneven distribution of the term sparsity in the model which results in idle lanes waiting for the slowest lane to finish execution (``no terms''), b)~high span across consecutive terms within a lane which cause stalls due to the limited per cycle shift range of the PE (``shift range'').

Stalls due to sharing the exponent block are rare. The more bit sparsity a model has the higher pressure on the exponent block and the higher the chance that it will not be able to keep up. Exponent stalls are noticeable for ResNet18-Q since the values there are effectively 4b. {We can see a similar behavior for SNLI due to its high bit sparisty.} However, there are other types of stalls that reduce pressure on the shared exponent block. First are stalls due to the limited per cycle shifting ability of the PEs. These are relatively few thus demonstrating that this technique presents a good performance vs. area trade-off. Second are stalls due to cross-lane term imbalance. These are the highest cause of PE underutilization;   32.8\% on average, and at most $55\%$ for NCF. Term imbalance is lowered if we reduce the number of lanes per PE or if we add weight buffers to allow faster lanes to proceed with the next set of weights albeit with a higher area overhead. However, doing so would increase the cost of the PE. This investigation is left for future work. Third are stalls due to inter-PE synchronization which are also rare. The ability for PE columns to run ahead by just one set sufficiently hides these stalls. \hl{Figure}~\ref{fig:sync_overhead_breakdown} \hl{shows the benefit of skipping out-of-bound terms in reducing the synchronization overheads (an average of $30.3\%$ overall reduction) by improving the load balancing across the PE lanes.}

\begin{figure*}[h!]
\begin{minipage}{.29\textwidth}
    \includegraphics[scale=0.24]{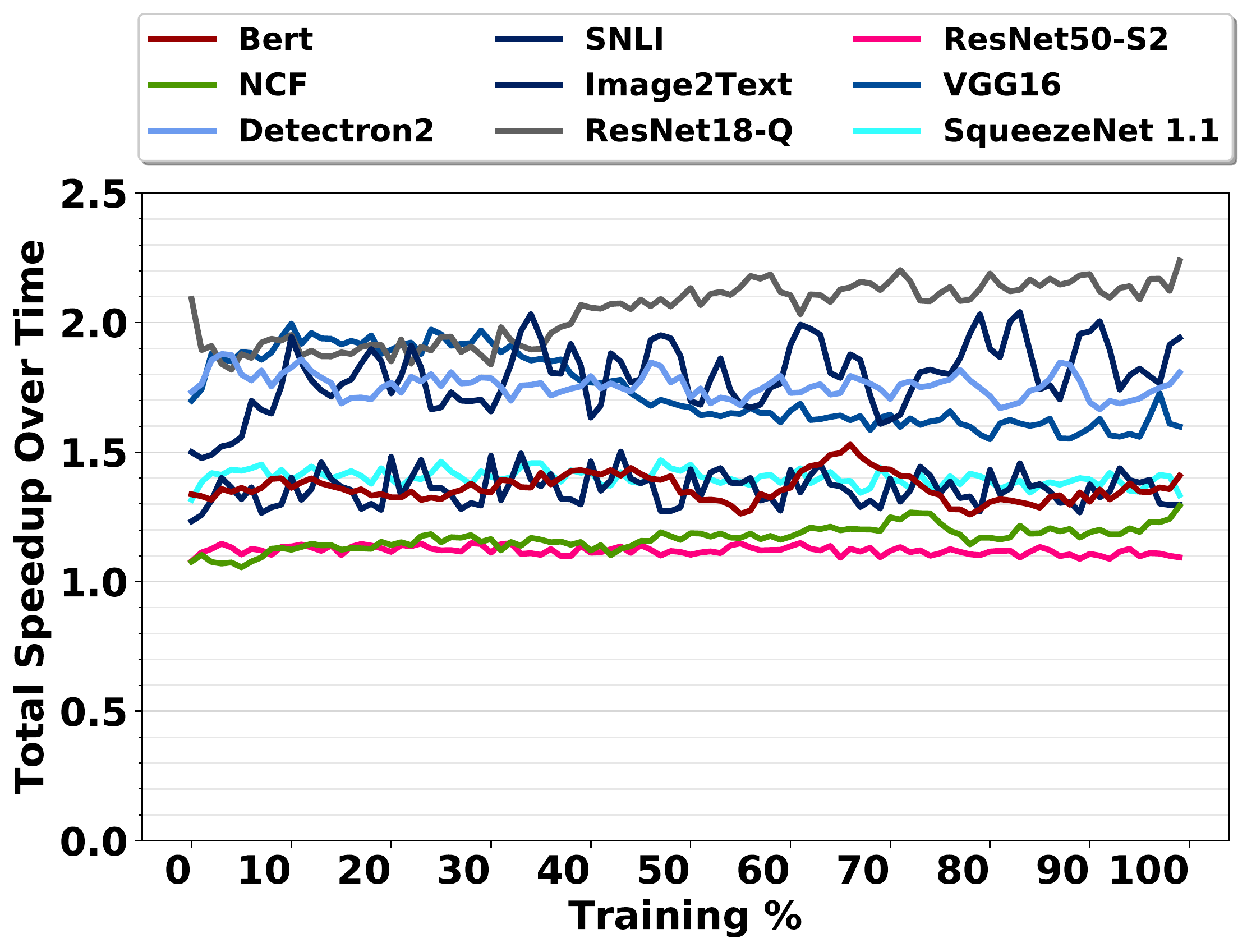}
    \vspace*{-0.7cm}
    \caption{Speedup of \OURS vs. the baseline over time.}
    \label{fig:speedup_over_time}
\end{minipage}
\hfill
\begin{minipage}{.29\textwidth}   
    \includegraphics[scale=0.23]{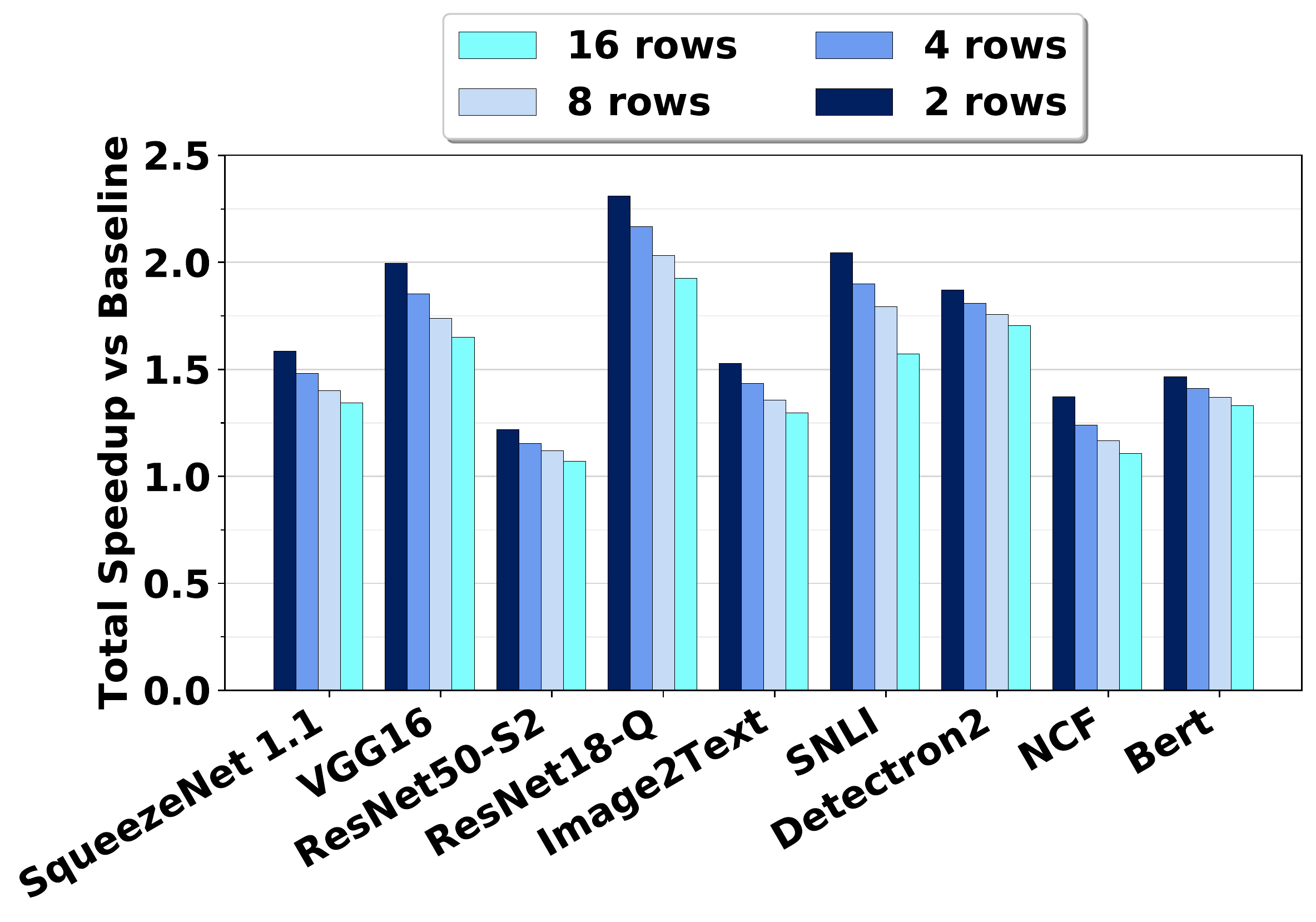}
    \vspace*{-0.7cm}
    \caption{Speedup of \OURS vs. the baseline with varying the number of rows per tile.}
    \label{fig:speedup_vary_row}
\end{minipage}
\hfill
\begin{minipage}{.35\textwidth}  
  \centering
  \includegraphics[scale=0.2]{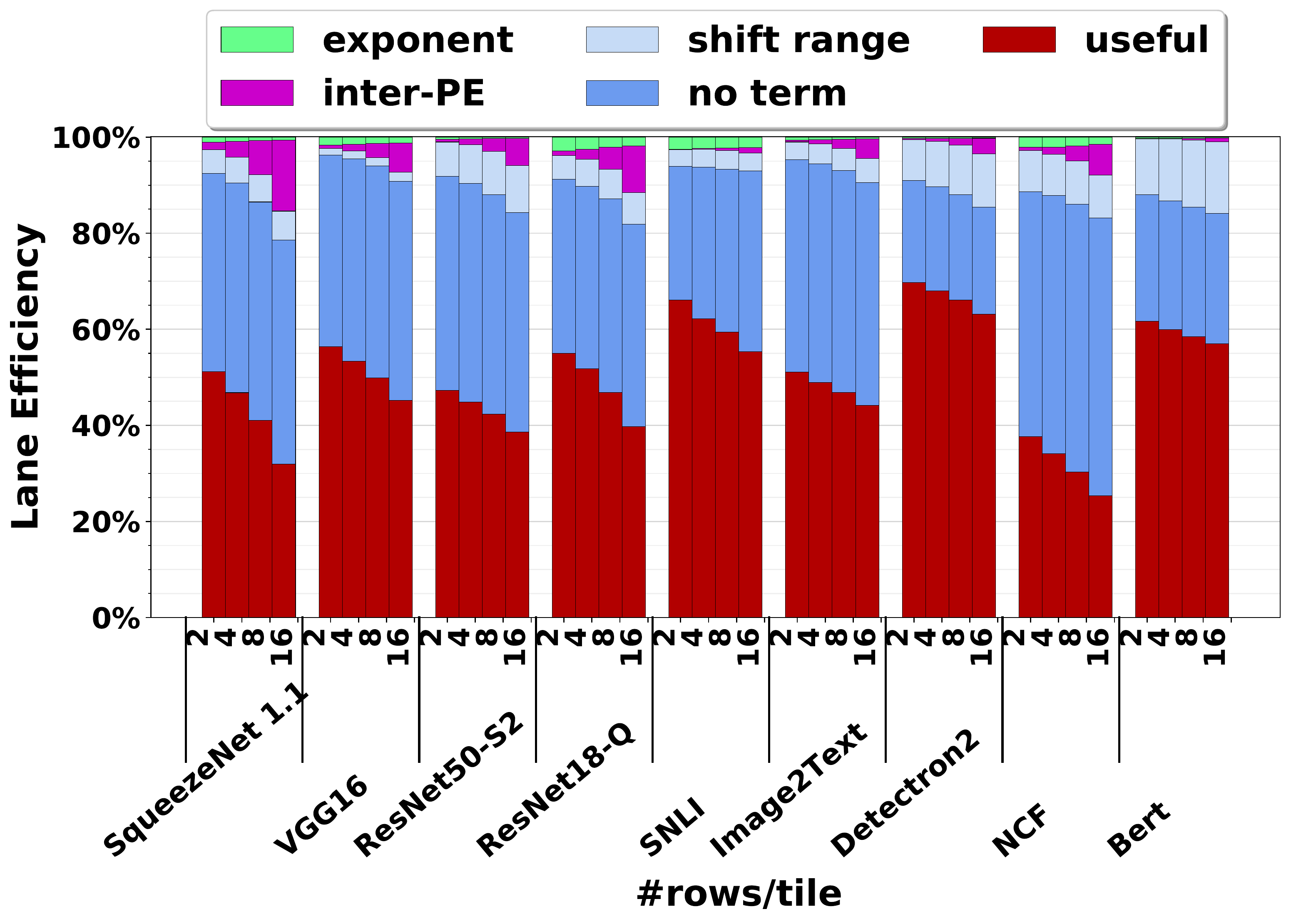}
  \vspace*{-0.7cm}
  \caption{Varying the number of rows: Breakdown of Cycles.}
  \label{fig:cycles_vary_row}
\end{minipage}
\vspace{-0.5cm}
\end{figure*}

\noindent{\textbf{Performance over Time: }}Figure~\ref{fig:speedup_over_time} shows the speedup of \OURS over the baseline over time and throughout the training process for all the studied networks. The measurements show three different trends. For VGG16 speedup is higher for the first 30 epochs after which it declines by around 15\% and plateaus. For ResNet18-Q, the speedup increases after epoch 30 by around 12.5\% and stabilizes. This can be attributed to the PACT clipping hyperparameter being optimized to quantize activations and weights within 4-bits or below. For the rest of the networks, speeds ups remain stable throughout the training process. Overall, the measurements show that performance of \OURS is robust and that it delivers performance improvements across all training epochs. 

\noindent{\textbf{Effect of Tile Organization: }}
As shown in Figure \ref{fig:speedup_vary_row}, increasing the number of rows per tile reduces performance by 6\% on average. This reduction in performance is due to synchronization among a larger number of PEs per column. When the number of rows increases, more PEs share the same set of A values. An A value that has more terms than the others will now affect a larger number of PE which will have to wait to finish processing. Since each PE processes a different combination of input vectors, each can be affected differently by intra-PE stalls such as ``no term'' stalls or ``limited shifting'' stalls. Figure~\ref{fig:cycles_vary_row} shows a breakdown of where time goes in each configuration. It can be seen that the stalls for the inter-PE synchronization increase and so do those for stalling for other lanes (``no term'').
\vspace*{-0.3cm}

\subsection{Accuracy Study}
\begin{figure}[H]
    \vspace{-0.4cm}
    \centering
    \includegraphics[width=0.9\linewidth]{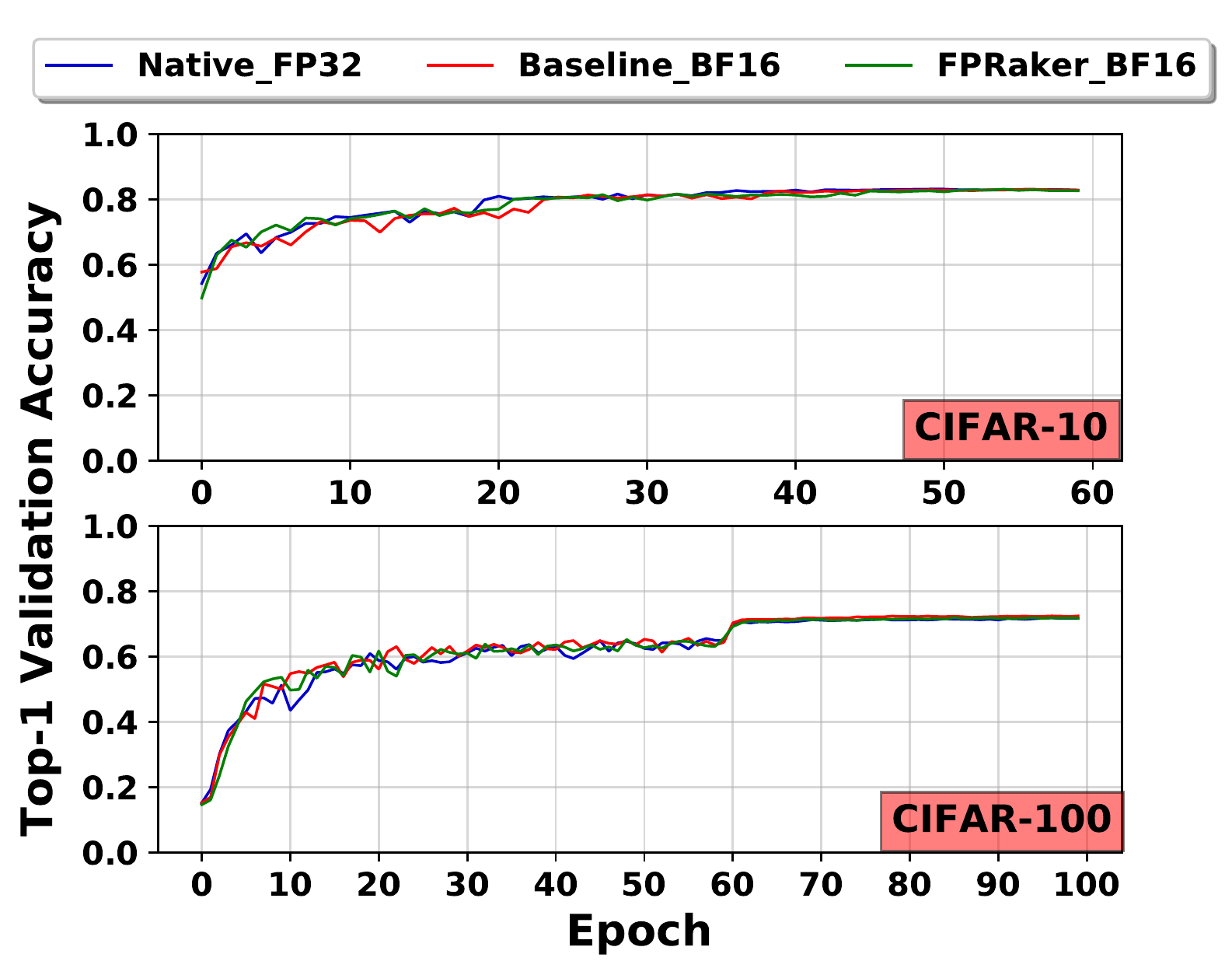} 
    \vspace{-0.4cm}
    \caption{Top-1 validation accuracy of training ResNet18 by emulating the FPRaker processing in PlaidML.}
    \label{fig:training}
    \vspace{-0.5cm}
\end{figure}

{To study the effect of training with FPRaker on accuracy, we emulated the bit-serial processing of FPRaker PE during end-to-end training in PlaidML~\cite{PlaidML} which is a machine learning framework based on an OpenCL compiler at the backend. We force PlaidML to use the mad() function for every multiply-add during training. We override the mad() function with our implementation to emulate the processing of the FPRaker PE. We trained ResNet18 on CIFAR-10 and CIFAR-100 datasets as shown in Fig.~\ref{fig:training}. The blue line shows the top-1 validation accuracy for training natively in PlaidML with FP32 precision. The baseline performs bit-parallel MAC with I/O operands precision in bfloat16 which is known to converge and supported in the industry, e.g. in Google's TPU. The figure shows that both the baseline and FPRaker emulated versions converge at epoch 60 for both datasets with accuracy difference within 0.1\% relative to the native training version. This is expected since FPRaker skips only ineffectual work, i.e., work which does not affect final result in the baseline MAC processing.}


\vspace*{-0.1cm}

\subsection{Per Layer Accumulator Width Profiling}
\begin{figure}[H]
  \vspace{-0.25cm}
  \centering
  \includegraphics[scale=0.27]{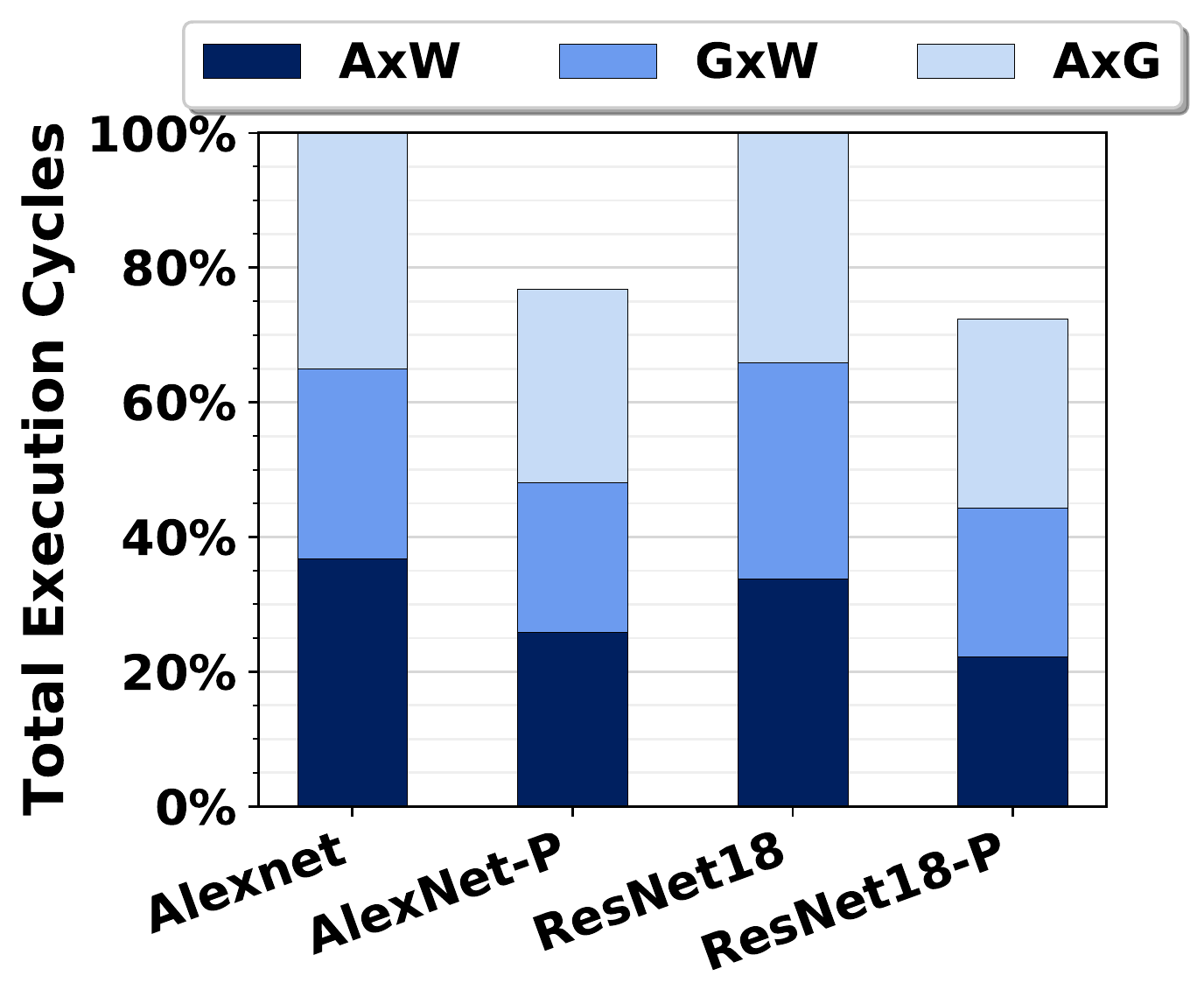}
  \vspace{-0.3cm}
  \caption{Performance of \OURS with per layer\\ profiled accumulator width~\cite{sakr2019accumulation} vs. fixed accumulator width.}
  \label{fig:perLayerAccWidth}
  \vspace{-0.3cm}
\end{figure}

Conventionally, training uses bfloat16 for all computations. As we noted in the introduction, there have been proposal for using mixed datatype arithmetic where some of the computations used fixed-point instead~\cite{mixedP, DBLP:conf/iclr/0002MMKAB0VKGHD18, Drumond:2018:TDH:3326943.3326985,  nvidia_mixedP}. Sakr et. al, propose to use floating-point where however the number of bits used by the mantissa varies per operation and per layer~\cite{sakr2019accumulation}. We use the suggested mantissa precisions while training AlexNet and ResNet18 on Imagenet. 
Figure~\ref{fig:perLayerAccWidth} shows the performance of \OURS following this approach. \OURS can dynamically take advantage of the variable accumulator width per layer to skip the ineffectual terms mapping outside the accumulator boosting overall performance. Training ResNet18 on ImageNet with per layer profiled accumulator width boosts the speedup of \OURS by $1.51\times$, $1.45\times$ and $1.22\times$ for $A\times W$, $G\times W$ and $A\times G$, respectively achieving an overall speedup of $1.56\times$ over the baseline compared to $1.13\times$ that is possiblew when training with a fixed accumulator width. Adjusting the mantissa length while using a bfloat16 container manifests itself a suffix of zero bits in the mantissa.

\vspace*{-0.5cm}

\section{Related Work}\label{sec:related}
We have discussed a broad spectrum of software level training acceleration techniques and some accelerator designs in the introduction. Since \OURL is a processing element level method it can in principle be used with any of these techniques to further boost performance and energy efficiency. However, there will be interactions which would need to be investigated. For example, since \OURL requires more parallelism to match and exceed the throughput of a conventional bit-parallel PE it will interact with data reuse/dataflow selection methods. However, data parallelism is abundant during training due to the use of batching. The investigation of these interactions while certainly interesting is left for future work. We believe that the experiments presented in this study sufficiently demonstrate that \OURL is a technique worthwhile considering further.

We have already commented on the conceptual differences between bit-Pragmatic and \OURL in the introduction which also apply to Laconic, which is another design that performs bit skipping for inference with fixed-point values. Further, we reported that converting bit-Pragmatic to process floating-point values results in a design that is less energy efficient compared to an optimized bit-parallel unit. We also converted Laconic to floating-point and unfortunately the results were equally disappointing. The design is not small enough nor energy efficient enough to improve over optimized bit-parallel floating point units.


{Compared to inference, training performs more computation and requires a larger volume of data to be kept around. While in inference activations and weights are used only once during training they are used twice and more importantly the dataflow is different between these two uses. This challenges several optimizations that are otherwise possible in inference. For example, inference accelerators targeting sparsity such as Cambricon-X}~\cite{CambriconXMICRO16} {or SCNN}~\cite{SCNN_ISCA} {prepack values in memory according to pre-determined dataflow to eliminate zero values. } {Unfortunately, this is not directly compatible with training as the order in which values will be accessed during backpropagation needs to be different than that during forward pass (inference).}
At the microarchitectural level, \OURL is a floating-point PE where the processing of values differs significantly from that of fixed-point values. The similarity between \OURL and Bit-Pragmatic is limited in the use of the 2-stage shifting technique which we adapted to reduce the area of our adder tree. Laconic uses a unique processing element which is designed for fixed-point arithmetic only. 

Bit-serial arithmetic circuits have long been used in embedded applications such as in signal processing with the emphasis being on reducing cost at the expense of performance. The closest related designs are the single multiplier of Shinde and Salankar~\cite{DBLP:conf/icacci/ShindeS15} and the single MAC of Chau et al.~\cite{Chau} which targets fault tolerance. Both use a multiplier stage, however, \OURL is a SIMD MAC that processes Booth terms across multiple value pairs eliminating the need for a multiplier stage and bares little similarity at the microarchitectural level with the aforementioned designs.

\hl{Feinberg et al.}~\cite{memristor_isca18}\hl{ proposes a memristive accelerator that supports double-precision floating-point \textit{in-situ} matrix-vector multiplication for scientific computing, e.g., krylov subspace methods.} \OURL,\hl{ however, is a purely digital DNN training accelerator supporting Bfloat16 precision and implemented using standard CMOS technology. Performing floating-point addition on a memristor crossbar requires converting the floating-point values into aligned fixed-point values increasing sparsity by extra padding bits. The memristive accelerator removes ineffectual computation on the crossbar-granularity by having different crossbar sizes per cluster and mapping blocks of effectual computations in the multiplicand matrix to similar-size crossbars through a pre-processing step. Inefficiently-mapped blocks are instead handled by the local processor.} \OURL,\hl{ however, performs regular floating-point addition and does not require fixed-point conversion.} \OURL\hl{ removes more ineffectual computations by skipping zero and out-of-bound terms per value through its term-serial processing.}

\section{Conclusion}\label{sec:conclude}
We presented \OURL, a novel processing element that performs multiple multiply-accumulate floating-point operations all contributing to a single final value. The processing element was developed as a building block for accelerators for training neural networks. We were motivated by the relatively high term level sparsity that all value exhibit during training. While we evaluated \OURL for training, it can naturally also be used for inference. While many neural network models can use fixed-point arithmetic there are models that still require floating-point. For example, these include models that process language or recommendation systems. Of course, whether \OURL will provide benefits for inference with such models needs to be demonstrated. 

{FPRaker opens new avenues of research in efficient precision training. Different precision can be assigned to each layer during training depending on the layer's sensitivity to quantization. Further, training can start with lower precision and increase the precision per epoch near conversion. FPRaker can adapt dynamically to different precisions and rewards innovations in training algorithms by boosting performance and energy efficiency.} Whether the benefits we demonstrate are compelling enough for practitioners to commit to deploying \OURL in the next generation designs can be of course be the subject of debate. Regardless, demonstrating this novel and practical approach to acceleration for floating-point values is of significant value to architects. We are confident that \OURL will spur additional work targeting computation and memory hierarchy optimizations that exploit term sparsity for improving training and inference.

\bibliographystyle{IEEEtran.bst}
\bibliography{ref}

\end{document}